\documentclass[ floatfix,%
 reprint,
 amsmath,amssymb,
 aps,
]{revtex4-2}

\usepackage{graphicx}
\usepackage{dcolumn}
\usepackage{bm}

\usepackage{multirow}
\usepackage{float}
\begin{document}

\preprint{APS/123-QED}

\title{Systematic Study of Proton, Two-Proton, Alpha, and Cluster Radioactivity Half-Lives based on the Deformed Gamow-like Model and Tabular Prior-data Fitted Network ($\mathrm{TabPFN}$)}

\author{Anqi Yang}
\author{Panpan Qi}
\author{Qingning Yuan}
\author{Gongming Yu}
\email{ygmanan@kmu.edu.cn}
\affiliation{College of Physics and Technology, Kunming University, Kunming 650214, China.}

\author{Haitao Yang}
\email{yanghaitao205@163.com}

\author{Zhangyan Li}
 \email{20230001@ztu.edu.cn}
\affiliation{%
School of Physics and Information Engineering, Zhaotong University, Zhaotong 657000, China
}%

\author{Yanbing Cai}
  \email{yanbingcai@mail.gufe.edu.cn}
\affiliation{%
Guizhou Key Laboratory in Physics and Related Areas, Guizhou University of Finance and Economics, Guiyang 550025, China
}%

\begin{abstract}
A hybrid framework that combines the deformed Gamow-like model (DGLM) with the Tabular Prior-data Fitted Network (TabPFN) has been developed to improve half-life predictions for two-proton emission, proton emission, $\alpha$ decay, and cluster radioactivity. A total of 583 radioactive nuclei are investigated, comprising 17 two-proton emitters, 42 proton emitters, 498 $\alpha$ emitters, and 26 cluster emitters. The DGLM+TabPFN framework achieves the best overall performance, with $\sigma_{\mathrm{RMS}}=0.423$, corresponding to an improvement of approximately $82.2\%$ over DGLM. The model parameters are optimized for each decay mode using the least-squares method. After incorporating TabPFN, the prediction errors for proton emission and $\alpha$ decay are reduced by approximately $80.6\%$ and $87.6\%$ respectively. For $\alpha$ decay, the training, test, and overall  $\sigma_{\mathrm{RMS}}$ values are 0.208, 0.305, and 0.240, respectively, indicating good generalization capability without apparent overfitting. The model also reproduces the systematic evolution of $\alpha$-decay half-lives and the shell-closure effect around $N=126$. These results demonstrate that combining DGLM with TabPFN significantly improves the accuracy and robustness of radioactive-decay half-life predictions while retaining the physical interpretability of the original model.
\end{abstract}

\maketitle
\enlargethispage{2\baselineskip}

\section{INTRODUCTION}

Radioactive nuclear decay is one of the fundamental topics in nuclear physics and provides important information on the structure, properties, and stability of atomic nuclei. A variety of theoretical models, empirical formulas, and computational approaches have been developed to predict the half-lives of various nuclear decay modes~\cite{Jinyu2026,xiao2026,Qi2026,Yuan2026,Ni2008,Qi2009,Santhosh2018}, including proton radioactivity~\cite{N.Teruya2016}, two-proton radioactivity~\cite{K.p.2022}, $\alpha$ decay~\cite{Zdeb.PRC2013, Zhao2026,Zhang2026,Jain2026,Qi2026EPJP,Jinyu2025,Jinyu2026_3,Zhang2007,Wan2021,Zhang2006,Wang2013}, and cluster radioactivity~\cite{Zdeb.PRC2013,Zhang2026,Warda2018,Jain2025,Majekodunmi2022}. Accurate predictions of their half-lives provide an important test of the reliability and applicability of theoretical decay models.

$\alpha$ decay is one of the most extensively studied radioactive-decay modes and has played an important role in the development of nuclear physics. In 1928, Gamow provided a quantum-mechanical explanation of $\alpha$ decay based on barrier penetration through quantum tunneling~\cite{Gamow1928}. At approximately the same time, Gurney and Condon independently developed a quantum-mechanical description of radioactive decay based on the same fundamental tunneling mechanism~\cite{Gurney1928}. These pioneering studies established the quantum-tunneling framework for radioactive decay and laid the foundation for subsequent theoretical investigations of nuclear decay processes. Since then, extensive theoretical and experimental studies of $\alpha$ decay have substantially deepened our understanding of nuclear structure and decay mechanisms. Proton radioactivity occurs in proton-rich nuclei near or beyond the proton drip line, where an emitted proton escapes through quantum tunneling across the Coulomb and centrifugal barriers. Two-proton ($2p$) radioactivity can occur in proton-rich nuclei when simultaneous two-proton emission is energetically allowed, whereas sequential one-proton emission is energetically forbidden for $Q_{2p}>0$ and $Q_p<0$. The decay is commonly treated as a three-body quantum system consisting of the daughter nucleus and the two emitted protons~\cite{Grigorenko2000,Grigorenko2001,Grigorenko2003,Grigorenko2007}. Cluster radioactivity is a spontaneous decay mode of heavy nuclei in which a nuclear cluster heavier than an $\alpha$ particle, such as $^{14}\mathrm{C}$, $^{20}\mathrm{O}$, and $^{24}\mathrm{Ne}$, is emitted~\cite{Ni2008,Ren2004,Zdeb.PRC2013}. Zdeb \textit{et al.}~\cite{Zdeb2013,Zdeb.PRC2013} developed a Gamow-like model based on the quantum-tunneling picture for describing $\alpha$ decay and cluster radioactivity. Owing to its simplicity and physical transparency, the Gamow-like approach has subsequently been widely applied and extended in radioactive-decay studies~\cite{Xing2022IMGL,Azeez2022}. 

In the original Gamow-like model, the nuclear interaction is represented by a square-well potential, while the external barrier is dominated by the Coulomb interaction. With the nuclear-radius constant $r_0$ as an adjustable parameter, the model can successfully reproduce experimental $\alpha$-decay half-lives for a wide range of even--even nuclei. An additional hindrance factor $h$ can be introduced to account for odd-mass and odd--odd nuclei~\cite{Ni2008,Ren2004}. The model therefore combines the simplicity of empirical approaches with the physical interpretation provided by the Gamow tunneling picture~\cite{Dong2009}. In 2025, Hollmann \textit{et al.} reported the Tabular Prior-data Fitted Network (TabPFN), a Transformer-based foundation model for tabular data~\cite{Noah2025}. TabPFN is particularly effective for prediction tasks involving relatively small tabular datasets and performs inference through an in-context learning framework. Its favorable performance on small datasets has motivated applications in nuclear-physics prediction problems~\cite{Zhao2026,Qi2026,Si_Yang2025,Yuan2026_2}.

In this work, we combine TabPFN with the deformed Gamow-like model (DGLM) to construct a hybrid framework for predicting the half-lives of proton radioactivity, two-proton radioactivity, $\alpha$ decay, and cluster radioactivity. The framework uses TabPFN to learn experimentally extracted preformation factors and subsequently incorporates the predicted preformation factors into the DGLM half-life calculations. The resulting predictions are compared with experimental data and with those obtained from established empirical or semi-empirical formulas to evaluate the accuracy and applicability of the proposed framework across different decay modes.

The remainder of this paper is organized as follows. Section~II introduces the theoretical framework of the DGLM--TabPFN approach, describes the four model configurations, and defines the root-mean-square deviation ($\sigma_{\mathrm{RMS}}$) used to evaluate their predictive performance. Section~III presents the results for two-proton radioactivity, proton radioactivity, $\alpha$ decay, and cluster radioactivity, with particular emphasis on the DGLM$^{b}$+TabPFN model. The nuclear-radius parameter $r_0$ and hindrance factor $h$ are optimized for the corresponding decay modes using the least-squares method, and the calculated half-lives are compared with experimental data and existing empirical or semi-empirical formulas. Finally, Section~IV summarizes the main conclusions of this work.

\section{THEORETICAL FRAMEWORK}

\subsection{Deformed Gamow-like Model}

In the DGLM framework, the half-lives $T_{1/2}$ of proton radioactivity, two-proton radioactivity, $\alpha$ decay, and cluster radioactivity can be written as 
\begin{equation}
T_{1/2}
=
\frac{\ln 2}{\lambda}\times 10^{h},
\label{eq:half_life}
\end{equation}
where $\lambda$ is decay constant, $h$ is an additional tunable parameter which is referred to as the hindrance factor for odd--even effects on the decay half-life~\cite{Zdeb2013}. For proton and two-proton radioactivity, as well as $\alpha$ decay in even--even nuclei and cluster radioactivity, $h$ is set to zero. In the present model, the adjustable parameters $r_0$ and $h$ are determined using the least-squares method.

The decay constant $\lambda$ is expressed as
\begin{equation}
\lambda = S\nu P,
\label{eq:lambda}
\end{equation}
where $S$, $\nu$, and $P$ are the preformation factor, assault frequency, and barrier penetration probability, respectively. Using Eqs.~\eqref{eq:half_life} and \eqref{eq:lambda}, together with the experimental half-life $T_{1/2}^{\mathrm{exp}}$, one obtains
\begin{equation}
\frac{T_{1/2}^{\mathrm{cal}}}
     {T_{1/2}^{\mathrm{exp}}}
=
\frac{S^{\mathrm{exp}}\nu P}
     {S^{\mathrm{cal}}\nu P}
=
\frac{S^{\mathrm{exp}}}
     {S^{\mathrm{cal}}},
\label{eq:inverse-relationship}
\end{equation}
where $S^{\mathrm{cal}}$ and $S^{\mathrm{exp}}$ are the Theoretical and experimentally extracted preformation factors, respectively. Following Ref.~\cite{Qi2026}, The experimental preformation factor $S^{\mathrm{exp}}$ can by obtained by setting the theoretical preformation factor $S^{\mathrm{cal}}$ to unity,
\begin{equation}
S^{\mathrm{exp}}
=
\frac{T_{1/2}^{\mathrm{cal}}}
     {T_{1/2}^{\mathrm{exp}}}.
\label{eq:S_exp}
\end{equation}

The assault frequency $\nu$ which is related to the harmonic-oscillator frequency in the Nilsson potential~\cite{Anyas-Weiss1974} can be written as
\begin{equation}
\nu=\frac{\omega}{2\pi},
\qquad
\hbar\omega
\approx
\frac{41}{A^{1/3}}~\mathrm{MeV},
\label{eq:nu}
\end{equation}
where $\hbar$, $\omega$, and $A$ are the reduced Planck constant, the angular frequency, and the mass number of the parent nucleus, respectively.

In the deformed Gamow-like model, the total barrier penetration probability $P$ is obtained by averaging the direction-dependent penetration probability over all orientations~\cite{Xu2006}:
\begin{equation}
P
=
\frac{1}{2}
\int_{0}^{\pi}
P(\theta)\sin\theta\,d\theta,
\label{eq:avg_penetration}
\end{equation}
where $\theta$ denotes the angle between the symmetry axis of the nucleus and the radial vector. Within the Wentzel--Kramers--Brillouin (WKB) approximation, the penetration probability along a given direction is expressed as
\begin{equation}
P(\theta)
=
\exp\left[
-\frac{2}{\hbar}
\int_{R_{\mathrm{in}}(\theta)}^{R_{\mathrm{out}}(\theta)}
\sqrt{2\mu\left|V(r,\theta)-E_k\right|}\,dr
\right],
\label{eq:wkb}
\end{equation}
where $\mu
=
\frac{m_em_d}{m_e+m_d}$ is the reduced mass, with $m_d$ and $m_e$ denoting the masses of thedaughter nucleus and the emitted particle, respectively. The kinetic energy of the emitted particle is given by
\begin{equation}
E_k
=
Q_e\frac{A-A_e}{A},
\end{equation}
where $Q_e$ is the decay energy, $A$ is the mass number of the parent nucleus, and $A_e$ is the mass number of the emitted particle. Here, $r$ is the distance between the centers of mass of the daughter nucleus and the emitted particle.

The inner turning point $R_{\mathrm{in}}(\theta)$ is the radius of the square potential well and can be expressed as
\begin{equation}
R_{\mathrm{in}}(\theta)
=
r_0A_e^{1/3}
+
r_0A_d^{1/3}
\left[
1+
\sum_{\lambda}
\beta_{\lambda}Y_{\lambda0}(\theta)
\right],
\label{eq:rin}
\end{equation}
where $r_0$ is the effective nuclear-radius constant and is treated as an adjustable parameter in the present model. $A_d$ and $A_e$ are the mass numbers of the daughter nucleus and the emitted particle, respectively. The outer turning point $R_{\mathrm{out}}$ satisfies
\begin{equation}
E_k=V(R_{\mathrm{out}}).
\end{equation}
The deformation parameters $\beta_{\lambda}$ with $\lambda=2$, $4$, and $6$ correspond to the quadrupole, hexadecapole, and hexacontatetrapole deformations, respectively. The corresponding deformation parameters can be found in the FRDM2012 model~\cite{P2016}, and $Y_{\lambda0}(\theta)$ is the spherical harmonic function.

The total interaction potential $V(r,\theta)$ between the emitted particle and the daughter nucleus is expressed as
\begin{equation}
V(r,\theta)
=
\begin{cases}
-V_0, & 0\leq r\leq R_{\mathrm{in}}(\theta),\\
V_C(r,\theta)+V_l(r), & r>R_{\mathrm{in}}(\theta),
\end{cases}
\label{eq:potential}
\end{equation}
where $V_0=25A_e~\mathrm{MeV}$~\cite{Blendowske1988} is the depth of the internal square-well potential, $V_C(r,\theta)$ is the Coulomb interaction between the daughter nucleus and the emitted particle. Considering the deformation of the daughter nucleus, the Coulomb potential is expressed as
\begin{equation}
\begin{split}
V_C(r,\theta)
&=
\frac{Z_dZ_ee^2}{r}
+
\frac{3Z_dZ_ee^2}{r}
\sum_{\lambda=2,4,6}
\frac{1}{2\lambda+1}
\\
&\quad\times
\left[
\frac{R_d(\theta)}{r}
\right]^{\lambda}
Y_{\lambda0}(\theta)\beta_{\lambda},
\end{split}
\label{eq:coulomb_potential}
\end{equation}
where $Z_d$ and $Z_e$ are the proton numbers of the daughter nucleus and the emitted particle, respectively. The centrifugal potential $V_l(r)$ is given by
\begin{equation}
V_l(r)
=
\frac{l(l+1)\hbar^2}{2\mu r^2},
\label{eq:centrifugal_potential}
\end{equation}
where $l$ is the orbital angular momentum of the emitted particle. The minimum value of $l$ is determined by parity and angular-momentum conservation~\cite{Qi2026}.

\subsection{Tabular Prior-Data Fitted Network Method}

Tabular Prior-Data Fitted Network (TabPFN) is a Transformer-based foundation model for tabular data, primarily suited for small- to medium-sized datasets with no more than $10\,000$ samples and $500$ features~\cite{Noah2025}. TabPFN is pretrained on a large synthetic tabular dataset and can used to make predictions on downstream datasets through in-context learning (ICL)~\cite{Si_Yang2025}. A dataset containing $N$ samples and $d$ input features can be represented as~\cite{Noah2025}
\begin{equation}
\mathcal{D}
=
\left\{
\left(
\mathbf{x}_i,y_i
\right)
\right\}_{i=1}^{N},
\label{eq:D}
\end{equation}
where $\mathbf{x}_i\in\mathbb{R}^{d}$ is the feature vector of the $i$-th sample, and $x_{i,j}$ denotes its $j$th feature. Input features may be continuous numerical variables or appropriately preprocessed categorical variables~\cite{Si_Yang2025}. For classification tasks, the labels satisfy $y_i\in\{1,\ldots,C\}$, where $C$ is the number of classes; for regression tasks, the targets are continuous,i.e., $y_i\in\mathbb{R}$. The training dataset is defined as~\cite{Noah2025}
\begin{equation}
\mathcal{D}_{\mathrm{train}}
=
\left\{
\left(
\mathbf{x}_{\mathrm{train}}^{(i)},
y_{\mathrm{train}}^{(i)}
\right)
\right\}_{i=1}^{N_{\mathrm{train}}},
\label{eq:Dtrain}
\end{equation}
where $\mathbf{x}_{\mathrm{train}}^{(i)}\in\mathbb{R}^{d}$ is the feature vector of the $i$th training sample, and $y_{\mathrm{train}}^{(i)}$ is its corresponding label. For a query sample $\mathbf{x}_{*}$, the TabPFN prediction can be expressed as
\begin{equation}
\widehat{y}_{*}
=
f
\left(
\mathbf{x}_{*}
\mid
\mathcal{D}_{\mathrm{train}}
\right),
\label{eq:tabpfn_prediction}
\end{equation}
where $f$ is the TabPFN prediction mapping determined by the pretrained parameters $\theta$, and $\widehat{y}_{*}$ denotes the predicted regression target~\cite{Noah2025}.

\subsection{DGLM--TabPFN Hybrid Framework}

Tabular data constitute one of the most widely used data formats across various application domains. TabPFN has recently emerged as a powerful foundation model for tabular prediction, offering strong predictive performance and efficient inference~\cite{Kyungeun2026}. By learning from a wide range of synthetic supervised-learning tasks during pretraining, TabPFN can perform predictions on downstream datasets through in-context learning without conventional task-specific parameter optimization~\cite{Kyungeun2026}.

Considering the preformation-factor regression, the nuclear-structure and decay-related quantities of each nuclide are combined to form the feature vector $\mathbf{x}_i$, while the logarithm of the experimentally extracted preformation factor can be treated as the continuous regression target~\cite{Qi2026}:
\begin{equation}
y_i
=
\log_{10} S_i^{\mathrm{exp}}.
\label{eq:y_i}
\end{equation}
For a query nuclide represented by $\mathbf{x}_{*}$, the corresponding TabPFN prediction in logarithmic space is expressed as
\begin{equation}
\widehat{y}_{*}
=
\log_{10} S_{\mathrm{pred},*}
=
f
\left(
\mathbf{x}_{*}
\mid
\mathcal{D}_{\mathrm{train}}
\right),
\label{eq:y_pred}
\end{equation}
where $\widehat{y}_{*}$ is the predicted target for the query nuclide. A smaller difference between $\widehat{y}_{*}$ and the corresponding ground-truth target $y_{*}$ indicates better predictive performance~\cite{Si_Yang2025}.

In this study, we develop a hybrid framework for predicting radioactive-decay half-lives by combining the DGLM with TabPFN. Instead of directly learning the experimental half-lives, TabPFN is used to learn the experimentally extracted preformation factors, which are subsequently incorporated into the DGLM to calculate the decay half-lives.

In the baseline DGLM, the theoretical preformation factor is initially set to $S^{\mathrm{cal}}=1$~\cite{Qi2026}. The decay half-lives are then calculated using the initial nuclear-radius constant $r_0^{a}=1.28~\mathrm{fm}$ and hindrance factor $h^{a}=0$, and this model is referred to DGLM$^{a}$. According to Eq.~\eqref{eq:S_exp}, the experimentally extracted preformation factor $S_{\mathrm{exp}}^{a}$ is obtained from the calculated and experimental half-lives.

The logarithm of $S_{\mathrm{exp}}^{a}$ is used as the regression target for TabPFN. The input feature vector consists of 12 physical quantities: the mass number $A$, proton number $Z$, neutron number $N$, decay energy $Q$, minimum orbital angular momentum $l$, deformation parameters $\beta_2$, $\beta_4$, and $\beta_6$, kinetic energy of the emitted particle $E_k$, internal square-well potential depth $V_0$, centrifugal potential $V_l$, and hindrance factor $h^{a}$. Through the in-context learning (ICL) mechanism of TabPFN, the preformation factor $S_{\mathrm{pred}}^{a}$ is predicted and then incorporated into DGLM$^{a}$ to calculate the corresponding decay half-life, and this hybrid model is referred to DGLM$^{a}$+TabPFN.

Furthermore, the nuclear-radius constant $r_0$ and hindrance factor $h$ are refitted for the different decay modes using the least-squares method. The optimized parameters, $r_0^{b}$ and $h^{b}$, is referred to the parameter-optimized DGLM$^{b}$. Based on the half-lives calculated with DGLM$^{b}$ and the corresponding experimental half-lives, the preformation factors $S_{\mathrm{exp}}^{b}$ are recalculated using Eq.~\eqref{eq:S_exp}. Following the same indirect-learning strategy, the logarithm of $S_{\mathrm{exp}}^{b}$ is used as the regression target, while the corresponding 12 physical quantities, with the optimized hindrance factor $h^{b}$ replacing $h^{a}$, are used as the input features. TabPFN then predicts the preformation factor $S_{\mathrm{pred}}^{b}$, which is incorporated into DGLM$^{b}$ to calculate the decay half-life, and this hybrid model is referred to DGLM$^{b}$+TabPFN.

Accordingly, four models are considered in this study: DGLM$^{a}$, DGLM$^{a}$+TabPFN, DGLM$^{b}$, and DGLM$^{b}$+TabPFN. To quantify their predictive performance, the root-mean-square deviation $\sigma_{\mathrm{RMS}}$ is defined as
\begin{equation}
\sigma_{\mathrm{RMS}}
=
\sqrt{
\frac{1}{n}
\sum_{i=1}^{n}
\left[
\log_{10}
\left(
T_{1/2,i}^{\mathrm{cal}}
\right)
-
\log_{10}
\left(
T_{1/2,i}^{\mathrm{exp}}
\right)
\right]^2
},
\label{eq:sigma_rms}
\end{equation}
where $n$ is the total number of decay data points, and $T_{1/2,i}^{\mathrm{cal}}$ and $T_{1/2,i}^{\mathrm{exp}}$ is the calculated and experimental half-lives of the $i$th data point, respectively.

\begin{table}[htbp]
\centering
\caption{Root-mean-square deviations $\sigma_{\mathrm{RMS}}$ obtained with the four models for the complete dataset of 583 decay data
points.}
\label{table:TabPFN}

\begin{tabular}{@{}l@{\hspace{0.5\columnwidth}}c@{}}
\hline
Model & $\sigma_{\mathrm{RMS}}$ \\
\hline
Number & 583 \\
DGLM$^{a}$ & 2.474 \\
DGLM$^{a}$+TabPFN & 0.516 \\
DGLM$^{b}$ & 2.383 \\
DGLM$^{b}$+TabPFN & 0.423 \\
\hline
\end{tabular}

\end{table}

The root-mean-square deviations $\sigma_{\mathrm{RMS}}$ between the half-lives predicted by the four models and the corresponding experimental values for the complete dataset of 583 decay data points are shown in Table~\ref{table:TabPFN}. The $\sigma_{\mathrm{RMS}}$ values obtained with DGLM$^{a}$, DGLM$^{a}$+TabPFN, DGLM$^{b}$, and DGLM$^{b}$+TabPFN are 2.474, 0.516, 2.383, and 0.423, respectively. The substantial decreases in $\sigma_{\mathrm{RMS}}$ after incorporating TabPFN indicate that learning the preformation factors considerably improves the agreement between the calculated and experimental half-lives. Among the four models, DGLM$^{b}$+TabPFN yields the smallest overall $\sigma_{\mathrm{RMS}}$, with a value of 0.423.

\begin{figure}
    \centering
    \includegraphics[width=0.85\linewidth]{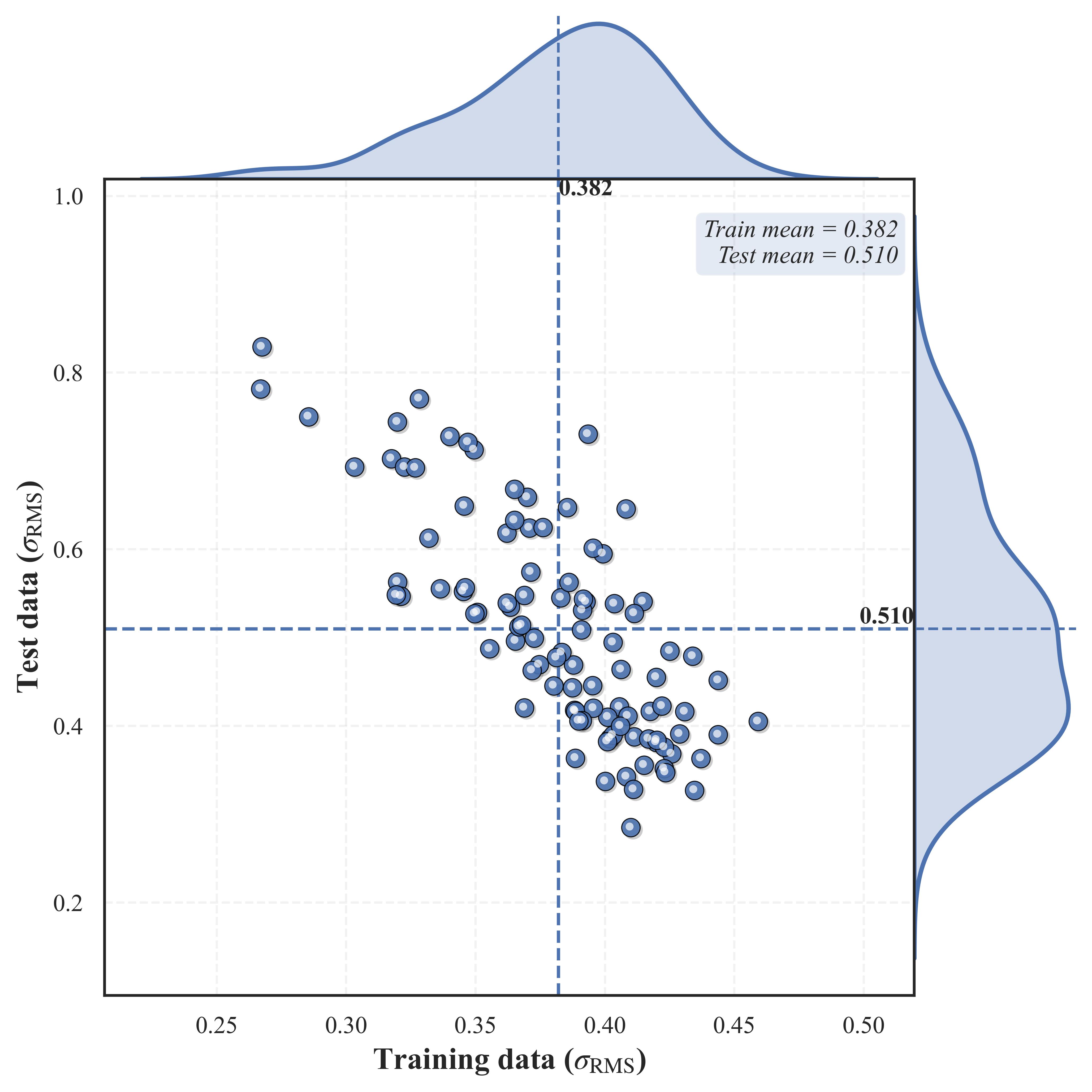}
    \caption{ Distribution of the $\sigma_{\mathrm{RMS}}$ values obtained from 100 independent random train--test splits of the dataset containing 583 decay data points. The dashed lines indicate the average $\sigma_{\mathrm{RMS}}$ values for the training and test sets, respectively.}
\label{fig:tabpfn}
\end{figure}

To reduce the dependence of the evaluation on a single random data split, DGLM$^{b}$+TabPFN is further evaluated over 100 independent train--test splits generated using different random seeds~ \cite{Zhao2026}. As shown in Fig.~\ref{fig:tabpfn}, the resulting $\sigma_{\mathrm{RMS}}$ values exhibit relatively limited variations across the different data partitions. The average $\sigma_{\mathrm{RMS}}$ values for the training and test sets over the 100 splits are 0.382 and 0.510, respectively. These results indicate that the predictive performance of the model is relatively stable across different random data partitions. Although the test-set $\sigma_{\mathrm{RMS}}$ is systematically larger than that of the training set, the relatively stable distributions across the 100 splits suggest that the reported performance is not determined by a single favorable train--test partition.

\subsection{GNL and UDL}

In this study, appropriate empirical or semi-empirical formulas are used to calculate the half-lives for different decay modes, with these two calculations included for comparison. For proton ($p$) radioactivity, the new Geiger--Nuttall law (GNL) proposed by Chen et al.~\cite{Chen2019} can be written as
\begin{equation}
\log_{10}T_{1/2}^{p}
=
a_p
\left(
Z_d^{0.8}+l
\right)
Q_p^{-1/2}
+b_p,
\label{eq:gnl_p}
\end{equation}
where the parameters are $a_p=0.843$ and $b_p=-27.194$, $T_{1/2}^{p}$ is the proton-radioactivity half-life, $Z_d$ is the proton number of the daughter nucleus, $l$ is the orbital angular momentum of the emitted proton, and $Q_p$ is the proton-decay energy. 

For two-proton ($2p$) radioactivity, the new GNL formula proposed by Liu et al.~\cite{Liu2021} is given by
\begin{equation}
\log_{10}T_{1/2}^{2p}
=
a_{2p}
\left(
Z_d^{0.8}+l^{\beta}
\right)
Q_{2p}^{-1/2}
+b_{2p},
\label{eq:gnl_2p}
\end{equation}
where $T_{1/2}^{2p}$ is the two-proton-radioactivity half-life, $Q_{2p}$ is the two-proton decay energy, and $\beta$ characterizes the contribution of the orbital angular momentum. The corresponding parameters are $a_{2p}=2.032$, $b_{2p}=-26.832$, and $\beta=0.25$.

For $\alpha$ decay and cluster radioactivity, the universal decay law (UDL) proposed by Qi et al.~\cite{Qi2009} can be written as
\begin{equation}
\log_{10}T_{1/2}^{\alpha/c}
=
a\chi'
+
b\rho'
+
c,
\label{eq:udl}
\end{equation}
where
$
\chi'
=
Z_c Z_d\sqrt{\frac{A}{Q_c}},
$
$
\rho'
=
\sqrt{
A Z_c Z_d
\left(
A_d^{1/3}+A_c^{1/3}
\right)
},
$
and
$
A
=
\frac{A_dA_c}{A_d+A_c}.
$
Here, $A_c$ and $Z_c$ are the mass number and proton number of the emitted particle or cluster, respectively, while $A_d$ and $Z_d$ are for the daughter nucleus. $Q_c$ is the corresponding decay energy. The UDL parameters for  $\alpha$ decay and cluster radioactivity are given by
\begin{equation}
a=0.3949,
\qquad
b=-0.3693,
\qquad
c=-23.7615.
\end{equation}

Furthermore, the GNL formulas given in Eqs.~\eqref{eq:gnl_p} and \eqref{eq:gnl_2p} are used to calculate the half-lives of proton and two-proton radioactivity, respectively, whereas the UDL in Eq.~\eqref{eq:udl} is used for both $\alpha$ decayMand cluster radioactivity.

\section{RESULTS AND DISCUSSION}

\subsection{Two-Proton Radioactivity}

\begin{figure}[htbp]
    \centering
    \includegraphics[width=0.85\linewidth]{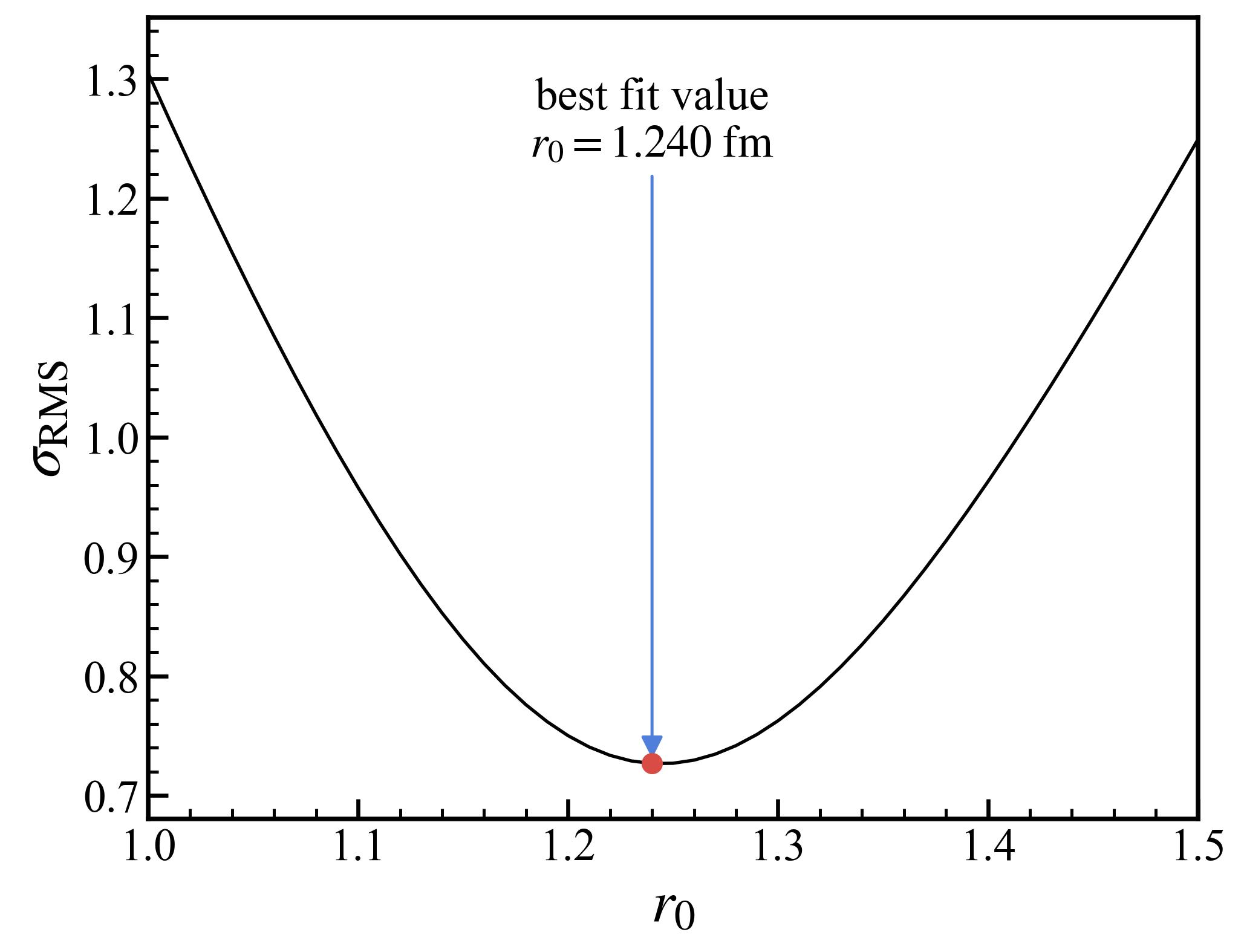}
    \caption{ Dependence of the root-mean-square deviation
    $\sigma_{\mathrm{RMS}}$ on the nuclear-radius parameter $r_0$
    for two-proton radioactivity.}
    \label{fig:2p_r0}
\end{figure}

As shown in Fig.~\ref{fig:2p_r0}, the adjustable nuclear-radius parameter $r_0$ in the deformed Gamow-like model for two-proton ($2p$) radioactivity is determined by the least-squares method. The optimized value is $r_0=1.24~\mathrm{fm}$, with the hindrance factor $h=0$~\cite{Zdeb2016}. To systematically investigate the half-lives of $2p$ radioactivity, we compare the calculated results obtained with DGLM$^{b}$ and DGLM$^{b}$+TabPFN with the corresponding experimental data and the predictions of the GNL formula, and the results are shown in Table~\ref{tab:2p_exp}. The first two columns list the $2p$ emitters and the corresponding decay energies $Q_{2p}$, respectively. The remaining four columns give the experimental half-lives and the calculated values obtained with DGLM$^{b}$+TabPFN, DGLM$^{b}$, and GNL, respectively.

For the $2p$ decay, the root-mean-square deviation obtained with DGLM$^{b}$ is $\sigma_{\mathrm{RMS}}=1.956$, whereas that obtained with DGLM$^{b}$+TabPFN decreases to $\sigma_{\mathrm{RMS}}=1.576$. Thus, incorporating TabPFN reduces the overall $\sigma_{\mathrm{RMS}}$ by approximately $19.4\%$ relative to DGLM$^{b}$, indicating an improvement in the overall agreement between the calculated and experimental $2p$-decay half-lives.

For short-lived $2p$ cases, such as ${}^{6}\mathrm{Be}$, ${}^{12}\mathrm{O}$, and ${}^{16}\mathrm{Ne}$, noticeable deviations between the calculated and experimental half-lives remain. In particular, for ${}^{16}\mathrm{Ne}$, the DGLM$^{b}$+TabPFN predictions differ from the corresponding experimental half-lives by more than two orders of magnitude for $Q_{2p}=1.33$ and $1.40~\mathrm{MeV}$. Previous studies have suggested that nuclear deformation and collective dynamics may affect the description of $2p$ radioactivity~\cite{Grigorenko2009,Rotureau2006}. 

\begin{table}[htbp]
\centering
\caption{Experimental and calculated half-lives of $2p$ radioactivity. The decay energies $Q_{2p}$ and experimental half-lives $T_{1/2}^{\mathrm{exp}}$ are taken from Ref.~\cite{Wang2026}.}
\label{tab:2p_exp}

\small
\setlength{\tabcolsep}{2.5pt}
\renewcommand{\arraystretch}{1.05}
\begin{tabular}{lccccc}
\hline
\multirow{2}{*}{${}^{A}\mathrm{Z}$} &
\multirow{2}{*}{$Q_{2p}$} &
\multicolumn{4}{c}{$\log_{10}T_{1/2}~(\mathrm{s})$} \\
\cline{3-6}
&  &Exp &
DGLM$^{b}$+TabPFN &
DGLM$^{b}$ &
GNL \\
\hline

${}^{6}\mathrm{Be}$  & 1.371 & $-20.30$ & $-19.90$ & $-21.59$ & $-23.81$ \\

${}^{12}\mathrm{O}$  & 1.638 & $-20.20$ & $-18.92$ & $-20.04$ & $-20.17$ \\
                      & 1.820 & $-20.94$ & $-19.32$ & $-20.30$ & $-20.52$ \\
                      & 1.790 & $-20.10$ & $-19.27$ & $-20.26$ & $-20.46$ \\
                      & 1.800 & $-20.12$ & $-19.29$ & $-20.28$ & $-20.48$ \\

${}^{16}\mathrm{Ne}$ & 1.330 & $-20.64$ & $-16.62$ & $-18.25$ & $-17.53$ \\
                      & 1.400 & $-20.38$ & $-16.94$ & $-18.45$ & $-17.77$ \\

${}^{19}\mathrm{Mg}$ & 0.750 & $-11.40$ & $-11.55$ & $-13.50$ & $-12.03$ \\

${}^{45}\mathrm{Fe}$ & 1.100 & $-2.40$ & $-2.20$ & $-4.19$ & $-2.21$ \\
                      & 1.140 & $-2.07$ & $-2.69$ & $-4.68$ & $-2.64$ \\
                      & 1.154 & $-2.55$ & $-2.85$ & $-4.84$ & $-2.79$ \\
                      & 1.210 & $-2.42$ & $-3.49$ & $-5.47$ & $-3.35$ \\

${}^{48}\mathrm{Ni}$ & 1.290 & $-2.52$ & $-2.65$ & $-4.68$ & $-2.59$ \\
                      & 1.350 & $-2.08$ & $-3.26$ & $-5.30$ & $-3.13$ \\

${}^{54}\mathrm{Zn}$ & 1.280 & $-2.79$ & $-1.63$ & $-3.02$ & $-1.01$ \\
                      & 1.480 & $-2.43$ & $-3.67$ & $-5.11$ & $-2.81$ \\

${}^{67}\mathrm{Kr}$ & 1.690 & $-1.70$ & $-1.52$ & $-2.90$ & $-0.58$ \\

\hline
\end{tabular}
\end{table}

Given the overall agreement between the results obtained with DGLM$^{b}$+TabPFN and the available experimental data, we further apply this model to predict the half-lives of potential $2p$ radioactive candidates. For comparison, the corresponding predictions obtained from the GNL formula are shown in Table~\ref{tab:2p_pre}. The first three columns give the candidate nuclides, the $2p$ decay energy $Q_{2p}$, and the orbital angular momentum $l$, respectively, while the last two columns give the half-lives predicted by DGLM$^{b}$+TabPFN and GNL.

\begin{table}[htbp] 
\centering \caption{Predicted half-lives of potential $2p$ radioactive candidates. The decay energies $Q_{2p}$ and orbital angular momenta $l$ are taken from Ref.~\cite{Gonçalves2017}.} 
\label{tab:2p_pre} 
\begin{tabular}{lcccc} 
\hline 
\multirow{2}{*}{${}^{A}\mathrm{Z}$} & 
\multirow{2}{*}{$Q_{2p}$} & 
\multirow{2}{*}{$\ell$} & 
\multicolumn{2}{c}{$\log_{10}T_{1/2}~(\mathrm{s})$} \\ 
\cline{4-5} & & & DGLM$^{b}$+TabPFN & GNL \\ \hline 
${}^{22}\mathrm{Si}$ & 1.283 & 0 & $-12.19$ & $-13.74$ \\ ${}^{26}\mathrm{S}$ & 1.755 & 0 & $-11.78$ & $-14.16$ \\ ${}^{34}\mathrm{Ca}$ & 1.474 & 0 & $-8.52$ & $-9.93$ \\ ${}^{36}\mathrm{Sc}$ & 1.993 & 0 & $-12.13$ & $-11.66$ \\ ${}^{38}\mathrm{Ti}$ & 2.743 & 0 & $-13.61$ & $-13.35$ \\ ${}^{39}\mathrm{Ti}$ & 0.758 & 0 & $-1.05$ & $-1.19$ \\ ${}^{40}\mathrm{V}$ & 1.842 & 0 & $-10.31$ & $-9.73$ \\ ${}^{42}\mathrm{Cr}$ & 1.002 & 0 & $-2.85$ & $-2.76$ \\ ${}^{47}\mathrm{Co}$ & 1.042 & 0 & $-0.66$ & $-0.69$ \\ ${}^{56}\mathrm{Ga}$ & 2.443 & 0 & $-9.08$ & $-7.61$ \\ ${}^{58}\mathrm{Ge}$ & 3.732 & 0 & $-12.09$ & $-10.85$ \\ ${}^{59}\mathrm{Ge}$ & 2.102 & 0 & $-7.16$ & $-5.54$ \\ ${}^{61}\mathrm{As}$ & 2.282 & 0 & $-0.50$ & $-5.85$ \\ ${}^{10}\mathrm{N}$ & 1.300 & 1 & $-17.29$ & $-18.59$ \\ ${}^{28}\mathrm{Cl}$ & 1.965 & 2 & $-11.60$ & $-12.46$ \\ ${}^{32}\mathrm{K}$ & 2.077 & 2 & $-11.19$ & $-11.55$ \\ ${}^{57}\mathrm{Ga}$ & 2.047 & 2 & $-6.56$ & $-4.14$ \\ ${}^{60}\mathrm{As}$ & 3.492 & 4 & $-9.22$ & $-8.33$ \\ 
\hline 
\end{tabular} 
\end{table}

\subsection{Proton Radioactivity}

\begin{figure}[htbp]
    \centering
    \includegraphics[width=0.85\linewidth]{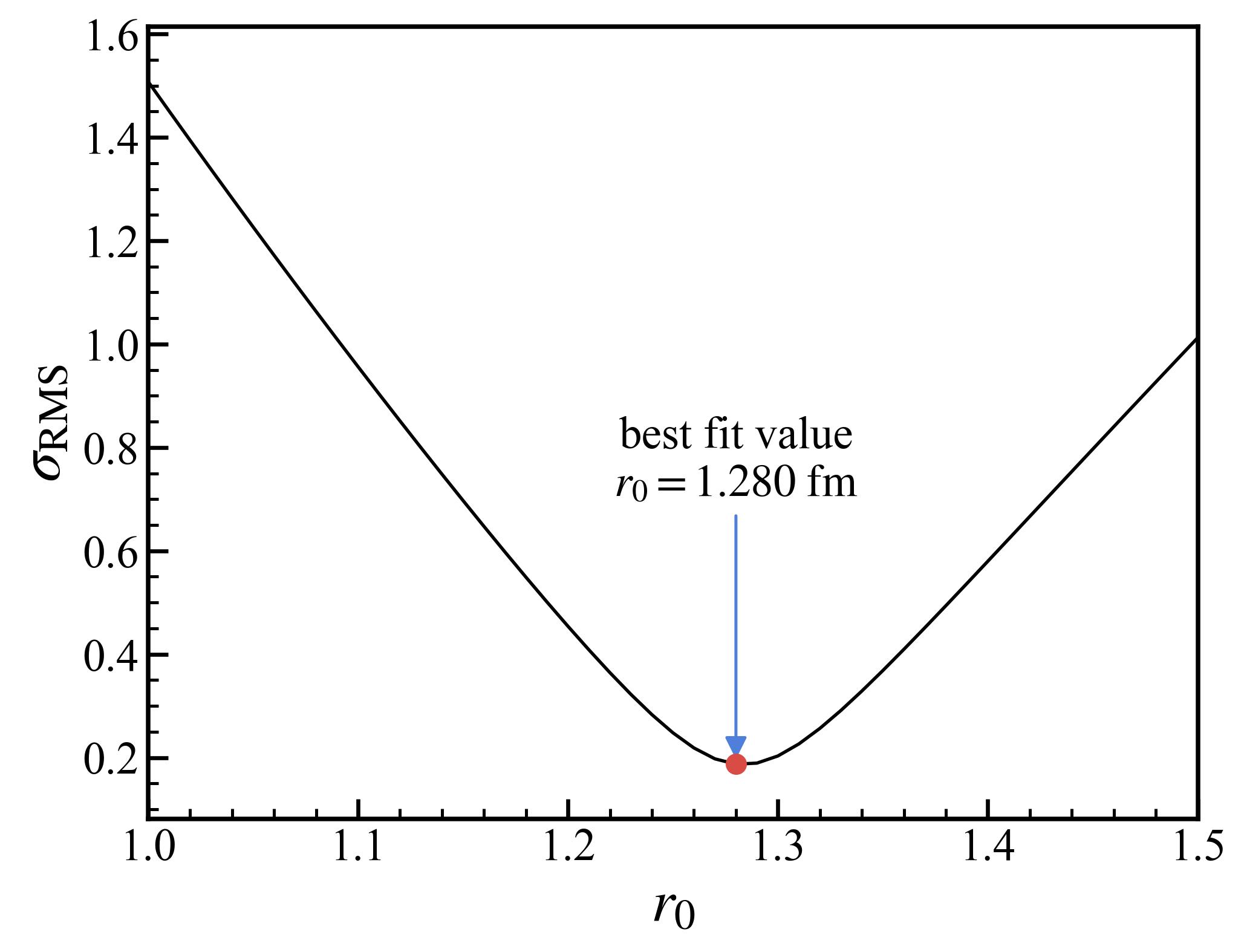}
    \caption{ Dependence of the root-mean-square deviation
    $\sigma_{\mathrm{RMS}}$ on the effective nuclear-radius parameter
    $r_0$ for proton radioactivity.}
    \label{fig:p_r0}
\end{figure}

For proton radioactivity, the effective nuclear-radius parameter $r_0$ in DGLM$^{b}$ is optimized using the least-squares method. As shown in Fig.~\ref{fig:p_r0}, the best agreement with the experimental data is obtained at $r_0=1.28~\mathrm{fm}$. Systematic calculations are performed for proton-emitting nuclei with proton numbers in the range $51\leq Z\leq83$, and the results are shown in Table~\ref{tab:p_exp}.
The comparison of experimental proton-radioactivity half-lives with the calculated results obtained using DGLM$^{b}$+TabPFN, DGLM$^{b}$, and the GNL formula are also shown in Table~\ref{tab:p_exp}. The first two columns list the proton-emitting nuclei and their corresponding decay energies $Q_p$, respectively, while the remaining four columns give the experimental and calculated values of $\log_{10}T_{1/2}$.
\begin{table}[htbp]
\centering
\caption{Experimental and calculated half-lives of proton emitters. The superscript $m$ is the first isomeric state. Experimental proton-emission half-lives, spins, and parities are taken from Ref.~\cite{Kondev2021}.The $Q_p$ values are obtained from Eq.\ref{eq:rin}, except for ${}^{130}\mathrm{Eu}$, ${}^{159}\mathrm{Re}$, ${}^{161}\mathrm{Re}^{m}$, ${}^{164}\mathrm{Ir}$, ${}^{177}\mathrm{Tl}^{m}$, and ${}^{185}\mathrm{Bi}$, for which the values are from Refs~\cite{Blank2008,Doherty2021}.}
\label{tab:p_exp}

\small
\setlength{\tabcolsep}{2.5pt}
\renewcommand{\arraystretch}{1.05}

\begin{tabular}{lccccc}
\hline
\multirow{2}{*}{${}^{A}\mathrm{Z}$} &
\multirow{2}{*}{$Q_p$} &
\multicolumn{4}{c}{$\log_{10}T_{1/2}~(\mathrm{s})$} \\
\cline{3-6}
&  &
Exp &
DGLM$^{b}$+TabPFN &
DGLM$^{b}$ &
GNL \\
\hline
${}^{109}\mathrm{I}$      & 0.829 & $-4.032$ & $-3.891$ & $-4.713$ & $-3.497$ \\
${}^{112}\mathrm{Cs}$     & 0.820 & $-3.310$ & $-2.911$ & $-3.864$ & $-2.694$ \\
${}^{113}\mathrm{Cs}$     & 0.981 & $-4.771$ & $-4.953$ & $-6.068$ & $-4.795$ \\
${}^{117}\mathrm{La}$     & 0.831 & $-1.664$ & $-1.914$ & $-3.397$ & $-2.193$ \\
${}^{121}\mathrm{Pr}$     & 0.901 & $-1.921$ & $-2.239$ & $-3.765$ & $-2.551$ \\
${}^{130}\mathrm{Eu}$     & 1.031 & $-3.000$ & $-2.759$ & $-4.226$ & $-2.985$ \\
${}^{131}\mathrm{Eu}$     & 0.963 & $-1.699$ & $-1.883$ & $-3.331$ & $-2.145$ \\
${}^{135}\mathrm{Tb}$     & 1.203 & $-2.996$ & $-3.342$ & $-4.822$ & $-3.477$ \\
${}^{140}\mathrm{Ho}$     & 1.104 & $-2.222$ & $-1.864$ & $-3.110$ & $-1.880$ \\
${}^{141}\mathrm{Ho}$     & 1.194 & $-2.387$ & $-2.903$ & $-4.140$ & $-2.852$ \\
${}^{141}\mathrm{Ho}^{m}$ & 1.264 & $-5.137$ & $-5.315$ & $-6.417$ & $-5.785$ \\
${}^{144}\mathrm{Tm}$     & 1.724 & $-5.569$ & $-5.168$ & $-5.978$ & $-5.209$ \\
${}^{145}\mathrm{Tm}$     & 1.754 & $-5.499$ & $-5.414$ & $-6.165$ & $-5.398$ \\
${}^{146}\mathrm{Tm}^{m}$ & 1.214 & $-1.137$ & $-0.932$ & $-1.655$ & $-0.995$ \\
${}^{147}\mathrm{Tm}$     & 1.072 & $0.587$  & $0.555$  & $0.090$  & $0.686$  \\
${}^{147}\mathrm{Tm}^{m}$ & 1.133 & $-3.444$ & $-3.271$ & $-3.605$ & $-2.451$ \\
${}^{150}\mathrm{Lu}$     & 1.285 & $-1.347$ & $-1.325$ & $-1.844$ & $-1.219$ \\
${}^{150}\mathrm{Lu}^{m}$ & 1.305 & $-4.398$ & $-4.506$ & $-4.902$ & $-3.633$ \\
${}^{151}\mathrm{Lu}$     & 1.255 & $-0.896$ & $-1.010$ & $-1.538$ & $-0.911$ \\
${}^{151}\mathrm{Lu}^{m}$ & 1.315 & $-4.796$ & $-4.613$ & $-5.012$ & $-3.723$ \\
${}^{155}\mathrm{Ta}$     & 1.466 & $-2.495$ & $-2.486$ & $-3.042$ & $-2.401$ \\
${}^{156}\mathrm{Ta}$     & 1.036 & $-0.826$ & $-0.745$ & $-1.049$ & $-0.185$ \\
${}^{156}\mathrm{Ta}^{m}$ & 1.126 & $0.933$  & $1.062$  & $0.576$  & $1.096$  \\
${}^{159}\mathrm{Re}$     & 1.816 & $-4.678$ & $-4.587$ & $-5.275$ & $-4.493$ \\
${}^{159}\mathrm{Re}^{m}$ & 1.816 & $-4.665$ & $-4.587$ & $-5.275$ & $-4.493$ \\
${}^{160}\mathrm{Re}$     & 1.276 & $-3.163$ & $-3.015$ & $-3.488$ & $-2.351$ \\
${}^{161}\mathrm{Re}$     & 1.216 & $-3.357$ & $-3.326$ & $-3.572$ & $-3.275$ \\
${}^{161}\mathrm{Re}^{m}$ & 1.338 & $-0.678$ & $-0.661$ & $-1.392$ & $-0.747$ \\
${}^{164}\mathrm{Ir}$     & 1.844 & $-3.947$ & $-4.085$ & $-5.081$ & $-4.247$ \\
${}^{165}\mathrm{Ir}^{m}$ & 1.727 & $-3.433$ & $-3.352$ & $-4.292$ & $-3.483$ \\
${}^{166}\mathrm{Ir}$     & 1.167 & $-0.824$ & $-0.892$ & $-1.643$ & $-0.691$ \\
${}^{166}\mathrm{Ir}^{m}$ & 1.347 & $-0.076$ & $-0.089$ & $-0.995$ & $-0.346$ \\
${}^{167}\mathrm{Ir}$     & 1.087 & $-1.120$ & $-1.060$ & $-1.308$ & $-1.350$ \\
${}^{167}\mathrm{Ir}^{m}$ & 1.262 & $0.842$  & $0.848$  & $-0.075$ & $0.544$  \\
${}^{170}\mathrm{Au}$     & 1.487 & $-3.487$ & $-3.571$ & $-4.555$ & $-3.251$ \\
${}^{170}\mathrm{Au}^{m}$ & 1.767 & $-2.971$ & $-3.011$ & $-4.171$ & $-3.327$ \\
${}^{171}\mathrm{Au}$     & 1.464 & $-4.652$ & $-4.668$ & $-5.068$ & $-4.457$ \\
${}^{171}\mathrm{Au}^{m}$ & 1.718 & $-2.587$ & $-2.617$ & $-3.821$ & $-2.989$ \\
${}^{176}\mathrm{Tl}$     & 1.278 & $-2.208$ & $-2.263$ & $-2.571$ & $-2.361$ \\
${}^{177}\mathrm{Tl}$     & 1.173 & $-1.174$ & $-0.991$ & $-1.258$ & $-1.273$ \\
${}^{177}\mathrm{Tl}^{m}$ & 1.967 & $-3.346$ & $-3.450$ & $-5.132$ & $-4.172$ \\
${}^{185}\mathrm{Bi}^{m}$ & 1.625 & $-4.191$ & $-4.303$ & $-5.505$ & $-4.732$ \\
\hline
\end{tabular}
\end{table}

\begin{figure}[htbp]
    \centering
    \includegraphics[width=\linewidth]{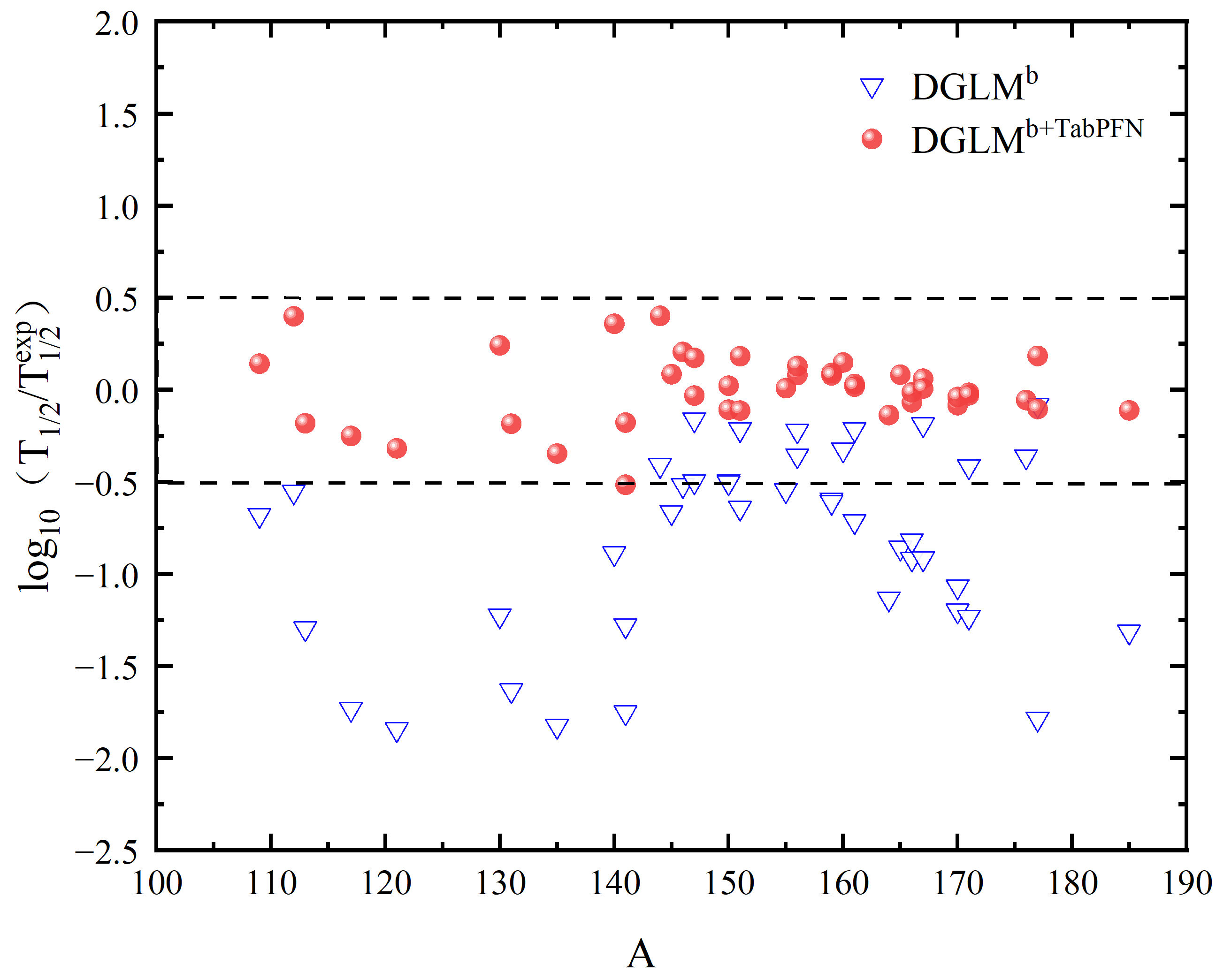}
    \caption{ Logarithmic deviations between the calculated and experimental proton-radioactivity half-lives as a function of the mass number $A$.}
    \label{fig:p_exp}
\end{figure}

To further compare the performance of the different models and quantify the improvement obtained by incorporating TabPFN into DGLM$^{b}$, we examine the logarithmic deviations between the calculated and experimental half-lives as a function of the mass number $A$, as shown in Fig.~\ref{fig:p_exp}. The results obtained from DGLM$^{b}$ generally exhibit negative deviations, with the deviations for several nuclei exceeding one order of magnitude, indicating a systematic tendency to underestimate the experimental half-lives. After incorporating TabPFN, the deviations obtained from DGLM$^{b}$+TabPFN are concentrated closer to zero, with most data points lying within $\pm0.5$. This behavior indicates that the incorporation of TabPFN substantially reduces the systematic deviations of DGLM$^{b}$ from the experimental data.

Quantitatively, the $\sigma_{\mathrm{RMS}}$ value decreases from 0.971 for DGLM$^{b}$ to 0.188 for DGLM$^{b}$+TabPFN, corresponding to a reduction of approximately $80.6\%$. This substantial reduction indicates that incorporating TabPFN considerably improves the overall agreement between the calculated and experimental proton-radioactivity half-lives.

Given the overall agreement between the DGLM$^{b}$+TabPFN results and the available experimental data for known proton emitters, we further apply this model to estimate the half-lives of six potential proton-emitting nuclei. These decay channels are energetically allowed or have been observed, but their proton-emission half-lives have not yet been quantified in NUBASE2020~\cite{Kondev2021}.  For comparison, the corresponding half-lives calculated using the GNL formula are also listed in Table~\ref{tab:p_pre}.
The first three columns of Table~\ref{tab:p_pre} list the candidate nuclei, proton-decay energies $Q_p$, and orbital angular momenta $l$, respectively, while the last two columns give the half-lives predicted by DGLM$^{b}$+TabPFN and GNL. The predictions of the two approaches are generally of comparable magnitude for these candidate nuclei, providing theoretical estimates that may be useful for future experimental studies of proton radioactivity.
\begin{table}[htbp]
\centering
\caption{Predicted half-lives of potential proton-radioactivity candidates. The decay energies $Q_p$ and orbital angular momenta $l$ are taken from Ref.~\cite{Kondev2021}.}
\label{tab:p_pre}
\small
\setlength{\tabcolsep}{3pt}
\renewcommand{\arraystretch}{1.05}

\begin{tabular}{lcccc}
\hline
\multirow{2}{*}{${}^{A}\mathrm{Z}$}     &
\multirow{2}{*}{$Q_p$}    &
\multirow{2}{*}{$\ell$}    &
\multicolumn{2}{c}{$\log_{10}T_{1/2}~(\mathrm{s})$} \\
\cline{4-5}
&  & &
DGLM$^{b}$+TabPFN &
GNL \\
\hline
${}^{103}\mathrm{Sb}$ & 0.979 & 2 & $-6.608$ & $-6.009$ \\
${}^{111}\mathrm{Cs}$ & 1.740 & 2 & $-10.548$ & $-10.375$ \\
${}^{116}\mathrm{La}$ & 1.591 & 2 & $-9.123$ & $-9.126$ \\
${}^{159}\mathrm{Re}$ & 1.606 & 0 & $-6.966$ & $-6.381$ \\
${}^{165}\mathrm{Ir}$ & 1.547 & 0 & $-5.960$ & $-5.530$ \\
${}^{169}\mathrm{Au}$ & 1.947 & 0 & $-8.125$ & $-7.478$ \\
\hline
\end{tabular}
\end{table}

\subsection{Alpha Decay}

\begin{figure*}[!t]
    \centering
    \includegraphics[width=0.85\textwidth]{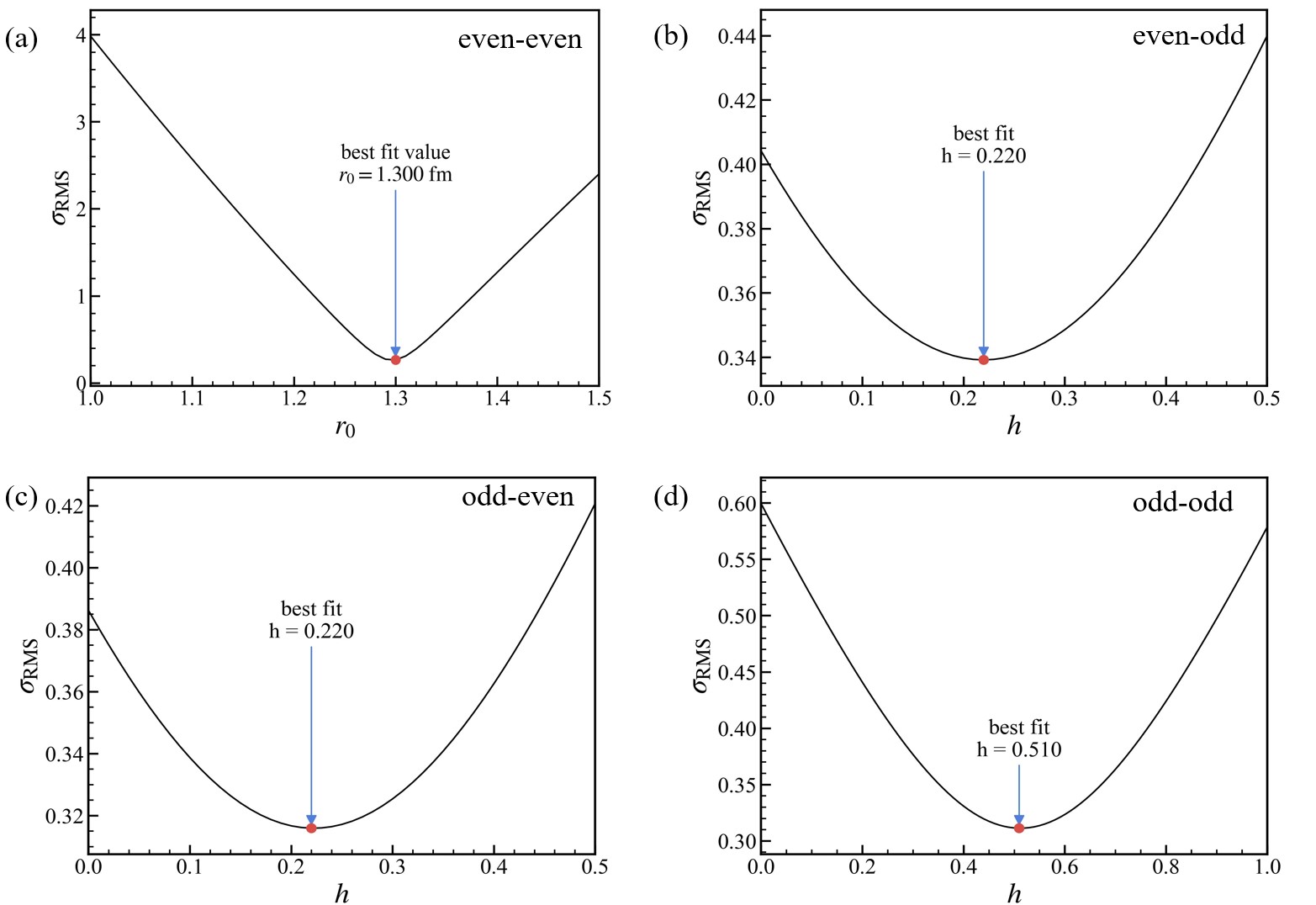}
    \caption{ Least-squares optimization of the $\alpha$-decay modelparameters $r_0$ and $h$:
    (a) $r_0$ for even--even nuclei;
    (b) $h$ for even-$Z$ odd-$N$ nuclei;
    (c) $h$ for odd-$Z$ even-$N$ nuclei; and
    (d) $h$ for odd--odd nuclei.
    The red dots indicate the minimum $\sigma_{\mathrm{RMS}}$ values and the corresponding
    optimal parameters.}
    \label{fig:alpha_r0_h}
\end{figure*}

For $\alpha$ decay, the effective nuclear-radius parameter $r_0$ and the hindrance factor $h$ are optimized using the least-squares method~\cite{Xiao2023}. The variable $r_0$ is determined from the experimental ground-state $\alpha$-decay data of 172 even--even nuclei by setting $h=0$. As shown in Fig.~\ref{fig:alpha_r0_h}(a), the $\sigma_{\mathrm{RMS}}$ reaches its minimum at $r_0=1.30~\mathrm{fm}$, and this value is therefore adopted in the subsequent calculations.

The hindrance factor $h$ is then determined separately using the experimental ground-state $\alpha$-decay data of 142 even-$Z$ odd-$N$, 119 odd-$Z$ even-$N$, and 65 odd--odd nuclei. As shown in Figs.~\ref{fig:alpha_r0_h}(b)--(d), the optimal values are
\begin{equation}
h_{\mathrm{EO}}=0.220,\qquad
h_{\mathrm{OE}}=0.220,\qquad
h_{\mathrm{OO}}=0.510.
\end{equation}
Each $\sigma_{\mathrm{RMS}}$ curve exhibits a well-defined minimum, indicating that the fitted parameters are reasonably stable. The even-$Z$ odd-$N$ and odd-$Z$ even-$N$ nuclei yield the same hindrance factor, $h=0.220$, where as the larger value obtained for odd--odd nuclei indicates a stronger hindrance effect in this nuclear category.
The experimental $\alpha$-decay half-lives, spins, and parities used in the present calculations are taken from NUBASE2016~\cite{Audi2017}. The $\alpha$-decay energies are taken from AME2016~\cite{Wang2021,Wang2017}, and the nuclear deformation parameters are taken from FRDM2012~\cite{P2016}.

Table~\ref{table:alpha_DGLMb_TabPFN_RMSE} summarizes the $\sigma_{\mathrm{RMS}}$ values of the DGLM$^{b}$+TabPFN model for the training set, test set, and entire dataset for different nuclear types. For all 498 nuclei, the $\sigma_{\mathrm{RMS}}$ values for the training set, test set, and entire dataset are 0.208, 0.305, and 0.240, respectively. Among the four nuclear types, even--even nuclei exhibit the lowest overall $\sigma_{\mathrm{RMS}}$ of 0.195. For odd--even nuclei, the test-set $\sigma_{\mathrm{RMS}}$ is 0.183, which is lower than the corresponding training-set value of 0.239. The test-set $\sigma_{\mathrm{RMS}}$ values for even--odd and odd--odd nuclei are 0.354 and 0.376, respectively. The relatively larger deviations for these nuclear types may reflect more complex nuclear-structure effects associated with unpaired nucleons. Overall, the results indicate that DGLM$^{b}$+TabPFN maintains reasonable predictive performance across different nuclear types.

\begin{table}[!t]
\centering
\caption{$\sigma_{\mathrm{RMS}}$ values of the DGLM$^{b}$+TabPFN model for different
nuclear types. }
\label{table:alpha_DGLMb_TabPFN_RMSE}

\small
\setlength{\tabcolsep}{4pt}
\renewcommand{\arraystretch}{1.05}

\begin{tabular}{lcccc}
\hline
Nuclear type & $n$ & Training set & Test set & Entire set \\
\hline
Even--even & 172 & 0.131 & 0.323 & 0.195 \\
Even--odd  & 142 & 0.260 & 0.354 & 0.284 \\
Odd--even  & 119 & 0.239 & 0.183 & 0.231 \\
Odd--odd   & 65  & 0.238 & 0.376 & 0.276 \\
All nuclei & 498 & 0.208 & 0.305 & 0.240 \\
\hline
\end{tabular}
\end{table}

\begin{figure}[!t]
    \centering
    \includegraphics[width=\linewidth]{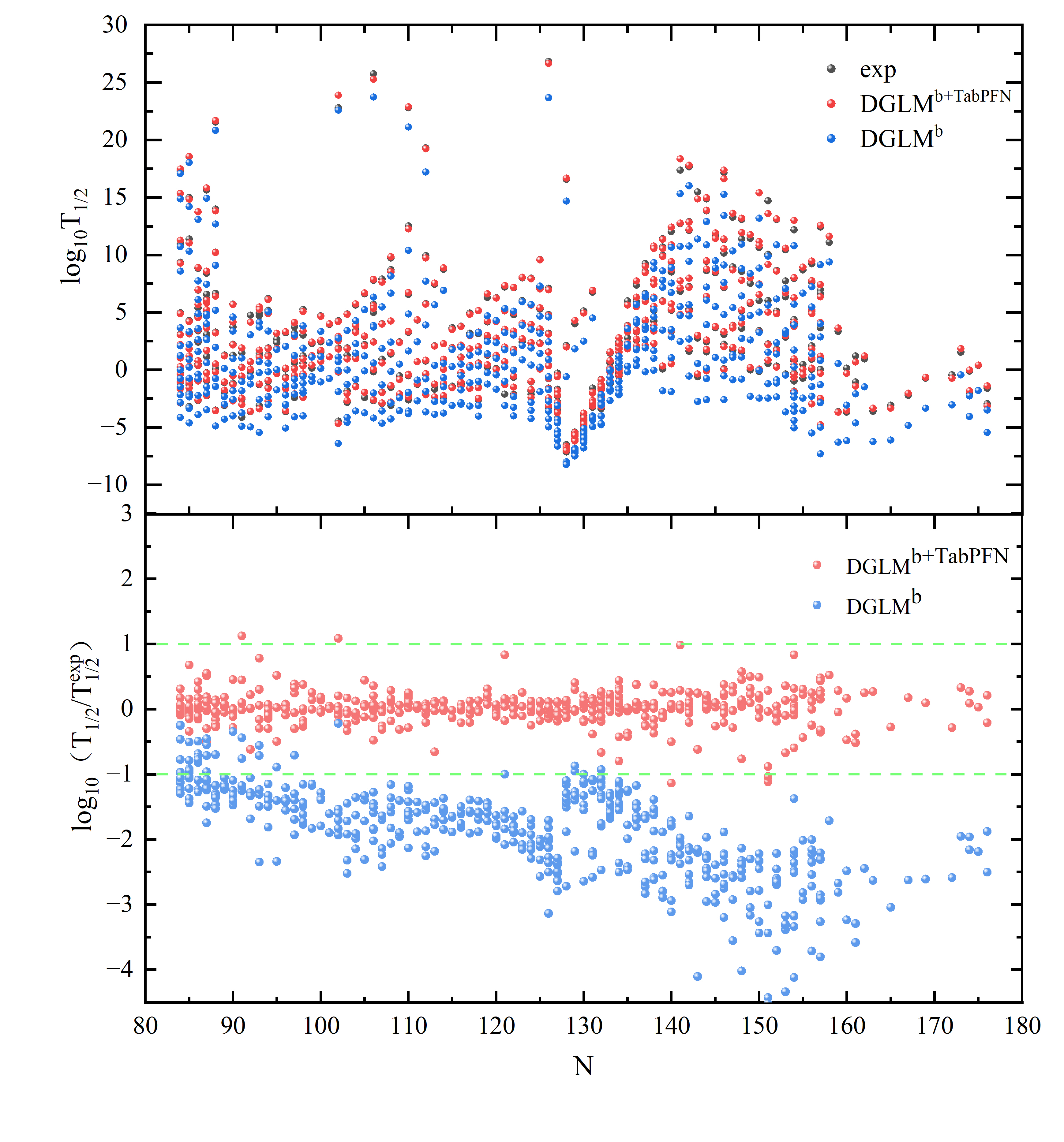}
    \caption{ Comparison of the calculated $\alpha$-decay half-lives
    with the experimental data for DGLM$^{b}$ and
    DGLM$^{b}$+TabPFN.
    (a) Calculated half-lives versus experimental values.
    (b) Logarithmic deviations
    $\log_{10}(T_{1/2}^{\mathrm{cal}}/T_{1/2}^{\mathrm{exp}})$.}
    \label{fig:alpha_exp}
\end{figure}

To further compare the performance of DGLM$^{b}$ and DGLM$^{b}$+TabPFN for $\alpha$ decay, Fig.~\ref{fig:alpha_exp} shows the calculated half-lives and their logarithmic deviations from the experimental values. Figure~\ref{fig:alpha_exp}(a) shows that the results obtained with DGLM$^{b}$+TabPFN are generally closer to the experimental values than those obtained with DGLM$^{b}$. As shown in Fig.~\ref{fig:alpha_exp}(b), DGLM$^{b}$ exhibits a systematic tendency to underestimate the experimental half-lives, whereas the deviations obtained with DGLM$^{b}$+TabPFN are more strongly concentrated around zero, with most values lying within $\pm1$. Quantitatively, the $\sigma_{\mathrm{RMS}}$ value decreases from 1.999 for DGLM$^{b}$ to 0.247 for DGLM$^{b}$+TabPFN, corresponding to a reduction of approximately $87.6\%$. This substantial reduction indicates that incorporating TabPFN considerably improves the overall agreement between the calculated and experimental $\alpha$-decay half-lives.

\begin{figure*}[!t]
    \centering
    \includegraphics[width=0.75\textwidth]{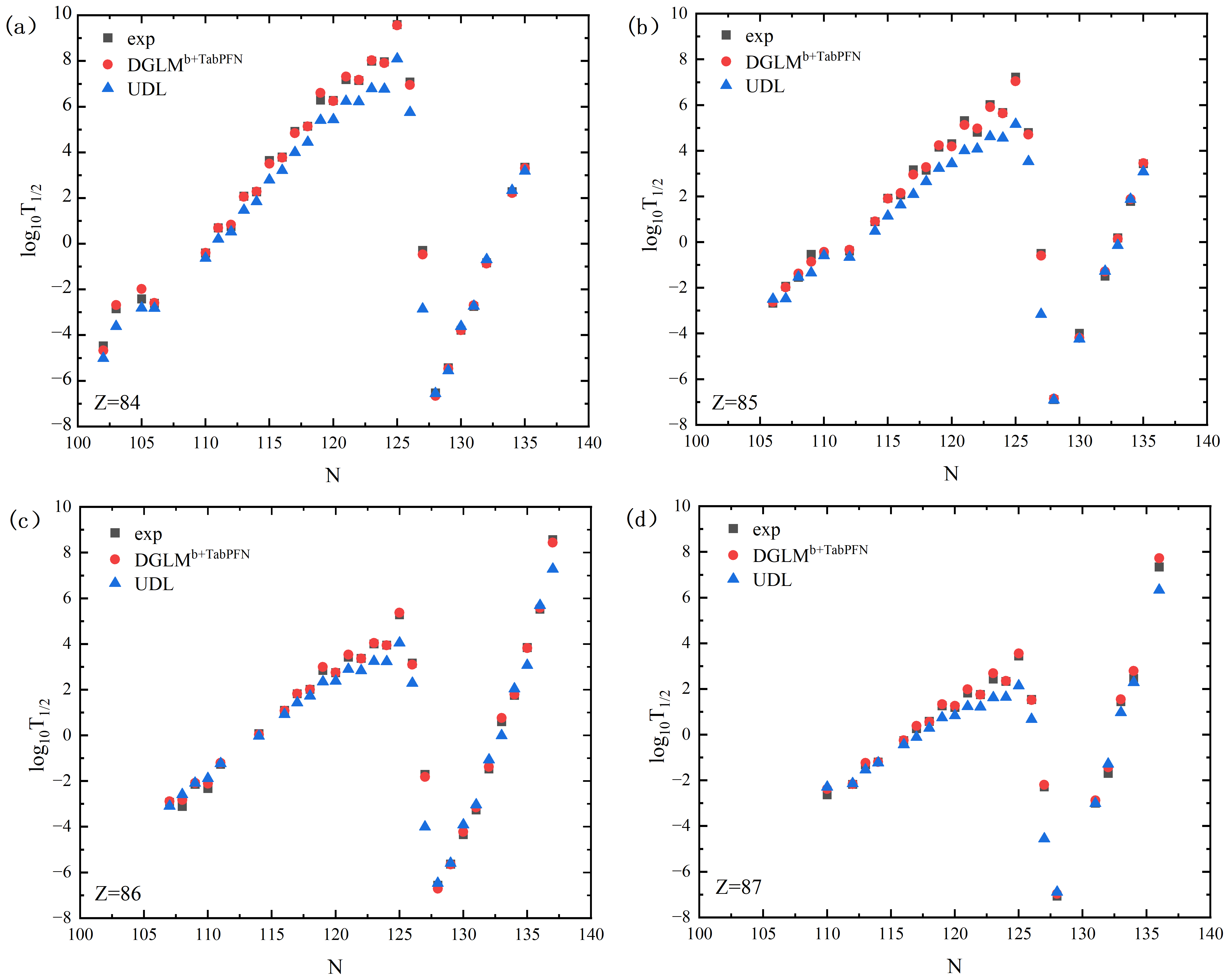}
    \caption{ Comparison of the experimental $\alpha$-decay half-lives with the predictions of DGLM$^{b}$+TabPFN and GNL for the
    (a) Po ($Z=84$), (b) At ($Z=85$), (c) Rn ($Z=86$), and
    (d) Fr ($Z=87$) isotopic chains as functions of neutron number
    $N$.}
    \label{fig:alpha_isotopic_chain}
\end{figure*}

To further examine the predictive performance of DGLM$^{b}$+TabPFN along different isotopic chains, Fig.~\ref{fig:alpha_isotopic_chain} compares the experimental $\alpha$-decay half-lives of the $Z=84$, 85, 86, and 87 isotopic chains with the results obtained using DGLM$^{b}$+TabPFN and the GNL formula. Both approaches reproduce the overall variation of the half-lives with neutron number $N$. For most of the nuclei considered, the DGLM$^{b}$+TabPFN results are closer to the experimental values. In particular, the pronounced variation of the experimental half-lives in the vicinity of the neutron shell closure at $N=126$ is also reproduced by the DGLM$^{b}$+TabPF calculations. These results indicate that the hybrid model provides an improved description of the systematic behavior of $\alpha$-decay half-lives along these isotopic chains.

To evaluate the predictive capability of DGLM$^{b}$+TabPFN for nuclei without experimentally determined $\alpha$-decay half-lives, we further perform calculations for selected nuclei shown in NUBASE2020~\cite{Kondev2021}. The $Q_{\alpha}$ values are from Ref.~\cite{Ma2019}, and the nuclear deformation parameters are taken from FRDM(2012)~\cite{P2016}. The predicted half-lives obtained from DGLM$^{b}$+TabPFN are compared with those calculated by the GNL formula, as shown in Table~\ref{tab:alpha_pre}.

\begin{table}[!t]
\centering
\caption{Predicted $\alpha$-decay half-lives of selected nuclei.}
\label{tab:alpha_pre}

\small
\setlength{\tabcolsep}{4pt}
\renewcommand{\arraystretch}{1.05}

\begin{tabular}{lcccc}
\hline
\multirow{2}{*}{${}^{A}\mathrm{Z}$} &
\multirow{2}{*}{$Q_{\alpha}$} &
\multirow{2}{*}{$\ell$} &
\multicolumn{2}{c}{$\log_{10}T_{1/2}~(\mathrm{s})$} \\
\cline{4-5}
& & &
DGLM$^{b}$+TabPFN &
GNL \\
\hline
${}^{266}\mathrm{Hs}$ & 10.35 & 0 & $-2.107$ & $-2.271$ \\
${}^{270}\mathrm{Hs}$ & 9.07  & 0 & $1.473$  & $1.195$  \\
${}^{290}\mathrm{Fl}$ & 9.86  & 0 & $0.775$  & $0.800$  \\
${}^{257}\mathrm{Rf}$ & 9.08  & 5 & $1.824$  & $-0.152$ \\
${}^{269}\mathrm{Sg}$ & 8.58  & 0 & $2.492$  & $1.980$  \\
${}^{263}\mathrm{Hs}$ & 10.73 & 5 & $-1.335$ & $-3.149$ \\
${}^{273}\mathrm{Hs}$ & 9.65  & 0 & $-0.067$ & $-0.532$ \\
${}^{275}\mathrm{Hs}$ & 9.45  & 0 & $0.441$  & $0.002$  \\
${}^{285}\mathrm{Fl}$ & 10.56 & 0 & $-0.758$ & $-1.015$ \\
${}^{287}\mathrm{Fl}$ & 10.17 & 0 & $0.211$  & $-0.016$ \\
${}^{291}\mathrm{Lv}$ & 10.89 & 0 & $-1.085$ & $-1.254$ \\
${}^{293}\mathrm{Lv}$ & 10.68 & 0 & $-0.674$ & $-0.752$ \\
${}^{271}\mathrm{Bh}$ & 9.42  & 0 & $0.271$  & $-0.222$ \\
${}^{275}\mathrm{Mt}$ & 10.48 & 0 & $-1.895$ & $-2.382$ \\
${}^{279}\mathrm{Rg}$ & 10.53 & 0 & $-1.484$ & $-1.880$ \\
${}^{285}\mathrm{Nh}$ & 10.01 & 0 & $0.324$  & $0.095$  \\
${}^{287}\mathrm{Mc}$ & 10.76 & 0 & $-0.976$ & $-1.209$ \\
${}^{289}\mathrm{Mc}$ & 10.49 & 0 & $-0.359$ & $-0.547$ \\
${}^{260}\mathrm{Bh}$ & 10.40 & 0 & $-2.025$ & $-2.650$ \\
${}^{262}\mathrm{Bh}$ & 10.32 & 0 & $-1.888$ & $-2.482$ \\
${}^{270}\mathrm{Bh}$ & 9.06  & 0 & $1.642$  & $0.848$  \\
${}^{272}\mathrm{Bh}$ & 9.30  & 0 & $0.874$  & $0.109$  \\
${}^{266}\mathrm{Mt}$ & 11.00 & 0 & $-2.848$ & $-3.489$ \\
${}^{276}\mathrm{Mt}$ & 10.10 & 0 & $-0.713$ & $-1.430$ \\
${}^{278}\mathrm{Mt}$ & 9.58  & 0 & $0.576$  & $-0.045$ \\
${}^{272}\mathrm{Rg}$ & 11.20 & 0 & $-2.548$ & $-3.377$ \\
${}^{274}\mathrm{Rg}$ & 11.48 & 0 & $-3.228$ & $-4.033$ \\
${}^{278}\mathrm{Rg}$ & 10.85 & 0 & $-1.899$ & $-2.648$ \\
${}^{280}\mathrm{Rg}$ & 10.15 & 0 & $-0.310$ & $-0.917$ \\
${}^{278}\mathrm{Nh}$ & 11.99 & 0 & $-3.733$ & $-4.552$ \\
${}^{282}\mathrm{Nh}$ & 10.78 & 0 & $-1.275$ & $-1.863$ \\
${}^{284}\mathrm{Nh}$ & 10.28 & 0 & $-0.079$ & $-0.618$ \\
${}^{286}\mathrm{Nh}$ & 9.79  & 0 & $1.200$  & $0.696$  \\
${}^{288}\mathrm{Mc}$ & 10.65 & 0 & $-0.434$ & $-0.945$ \\
${}^{290}\mathrm{Mc}$ & 10.41 & 0 & $0.113$  & $-0.351$ \\
\hline
\end{tabular}
\end{table}

\begin{figure*}[!t]
    \centering
    \includegraphics[width=0.75\textwidth]{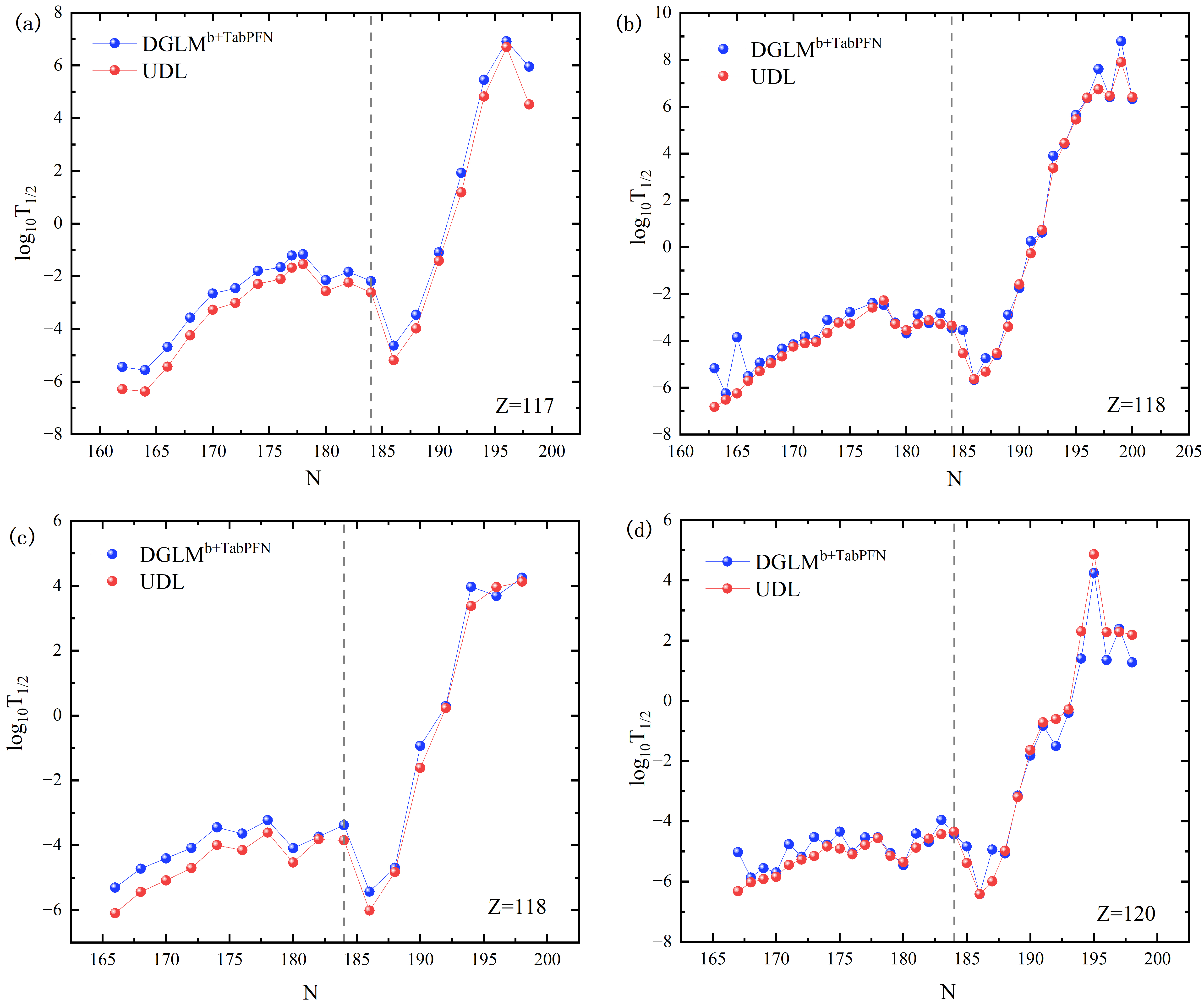}
    \caption{ Comparison of the predicted $\alpha$-decay half-lives
    $\log_{10}T_{1/2}$ obtained with DGLM$^{b}$+TabPFN and GNL for the superheavy isotopic chains
    (a) $Z=117$, (b) $Z=118$, (c) $Z=119$, and (d) $Z=120$.
    The vertical dashed line denotes $N=184$.}
    \label{fig:alpha_pre}
\end{figure*}

The DGLM$^{b}$+TabPFN and GNL predictions exhibit broadly similar trends for the selected nuclei in Table~\ref{tab:alpha_pre}. To further examine the extrapolation behavior of the hybrid model in the superheavy region, DGLM$^{b}$+TabPFN is subsequently applied to the $Z=117$--$120$ isotopic chains. The corresponding predictions are compared with those obtained using the GNL formula in Fig.~\ref{fig:alpha_pre}.

As shown in Fig.~\ref{fig:alpha_pre}, DGLM$^{b}$+TabPFN and GNL predict similar overall variations of the $\alpha$-decay half-lives with neutron number for the $Z=117$--$120$ isotopic chains. In particular, both calculations exhibit a pronounced change in the predicted half-lives around $N=184$~\cite{Qi2026}. This behavior is consistent with the expected influence of the neutron shell closure at $N=184$. Nevertheless, because experimental $\alpha$-decay half-lives are not available for these extremely neutron-rich superheavy nuclei, the predicted behavior should be regarded as a theoretical extrapolation rather than direct evidence for a shell closure. Differences between the two approaches also remain for several nuclei, reflecting the uncertainty associated with extrapolations far from the experimentally known region.

\subsection{Cluster Radioactivity}

For cluster radioactivity, the effective nuclear-radius parameter $r_0$ and the hindrance factor $h$ are optimized using the least-squares method. Since no experimentally confirmed odd--odd cluster emitters are included in the present dataset, only even--even and odd-$A$ cluster emitters are considered in the parameter optimization~\cite{Zdeb.PRC2013}.

\begin{figure*}[!t]
    \centering
    \includegraphics[width=0.8\textwidth]{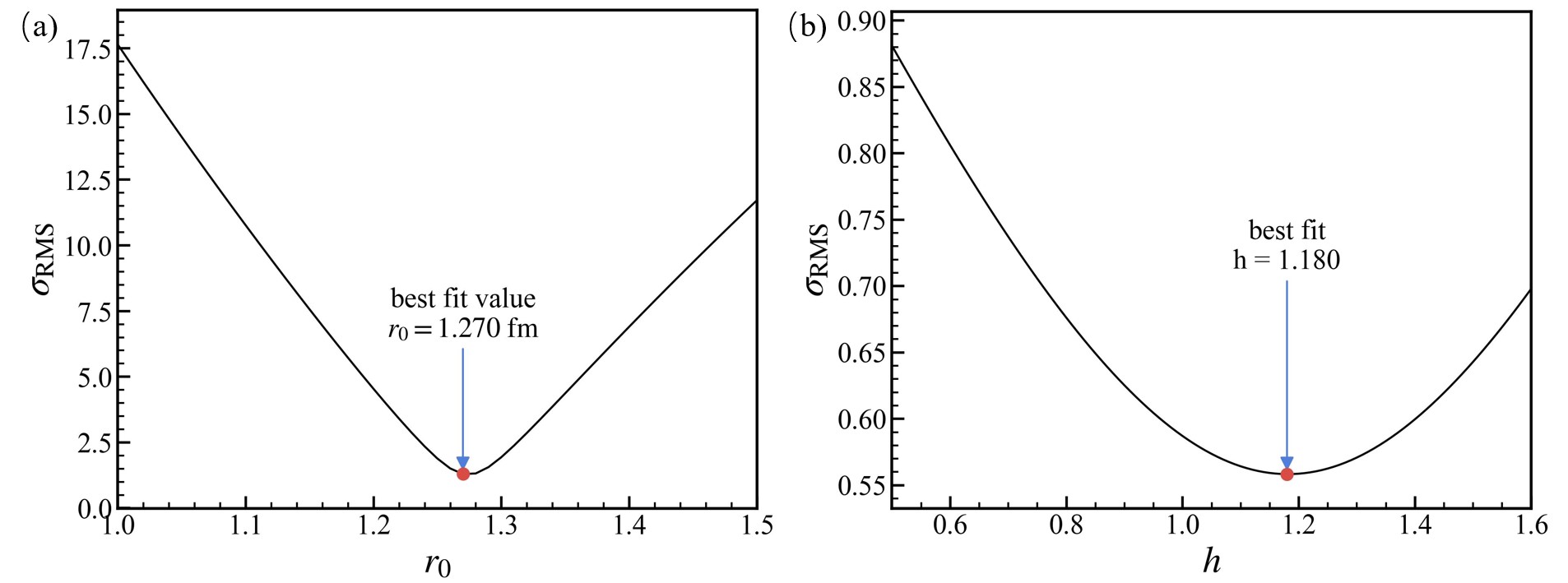}
    \caption{ Least-squares optimization of the cluster-radioactivity model parameters $r_0$ and $h$:
    (a) $r_0$ for even--even nuclei and
    (b) $h$ for odd-$A$ nuclei.
    The red dots indicate the minimum $\sigma_{\mathrm{RMS}}$ values and the corresponding
    optimal parameters.}
    \label{fig:c_r0_h}
\end{figure*}

Indeed, the nuclear-radius parameter $r_0$ is determined from the experimental ground-state cluster-decay half-lives of even--even nuclei by setting $h=0$. The optimal value is obtained by minimizing the root-mean-square error ($\sigma_{\mathrm{RMS}}$) between the calculated and experimental logarithmic half-lives. As shown in Fig.~\ref{fig:c_r0_h}(a), the $\sigma_{\mathrm{RMS}}$ exhibits a well-defined minimum at $r_0=1.27~\mathrm{fm}$. Therefore, $r_0=1.27~\mathrm{fm}$ is adopted in the subsequent calculations.

The hindrance factor $h$ is then determined from the experimental ground-state cluster-decay half-lives of odd-$A$ nuclei. As shown in Fig.~\ref{fig:c_r0_h}(b), the $\sigma_{\mathrm{RMS}}$ reaches its minimum at $h=1.18$. This hindrance factor phenomenologically accounts for the additional hindrance associated with the presence of an unpaired nucleon in odd-$A$ nuclei. The optimized values of $r_0$ and $h$ are used consistently in the DGLM$^{b}$ and DGLM$^{b}$+TabPFN calculations.
To evaluate the performance of the hybrid model for cluster radioactivity, calculations are performed for 26 experimentally known cluster-decay cases in the trans-lead region. The calculated half-lives are compared with the experimental data and with the results obtained using DGLM$^{b}$ and GNL, as summarized in Table~\ref{tab:cluster_exp}. The parent nuclei and decay energies $Q_{\mathrm{c}}$ are taken from NUBASE2020~\cite{Kondev2021} and AME2020~\cite{Wang2021}, respectively, while the experimental cluster-decay half-lives are taken from Refs.~\cite{Liu2024,Zhang2009,Warda2011,Saidi2015}.

\small
\setlength{\tabcolsep}{2.5pt}
\renewcommand{\arraystretch}{1.05}
\begin{table}[htbp]
\centering
\caption{Experimental and calculated cluster-radioactivity half-lives for selected trans-lead nuclei.}
\label{tab:cluster_exp}

\small
\setlength{\tabcolsep}{3.0pt}
\renewcommand{\arraystretch}{1.05}

\begin{tabular}{cccccc}
\hline
\multirow{2}{*}{Parent} &
\multirow{2}{*}{Cluster} &
\multirow{2}{*}{$Q_{\mathrm{c}}$} &
\multicolumn{3}{c}{$\log_{10}T_{1/2}~(\mathrm{s})$} \\
\cline{4-6}
& & &
Exp&
DGLM$^{b}$+TabPFN &
GNL \\
\hline
${}^{221}\mathrm{Fr}$ & ${}^{14}\mathrm{C}$  & 31.29 & 14.56 & 14.69 & 15.42 \\
${}^{221}\mathrm{Ra}$ & ${}^{14}\mathrm{C}$  & 32.40 & 13.39 & 13.49 & 14.25 \\
${}^{222}\mathrm{Ra}$ & ${}^{14}\mathrm{C}$  & 33.05 & 11.00 & 11.09 & 12.99 \\
${}^{223}\mathrm{Ra}$ & ${}^{14}\mathrm{C}$  & 31.83 & 15.05 & 14.53 & 15.27 \\
${}^{224}\mathrm{Ra}$ & ${}^{14}\mathrm{C}$  & 30.53 & 15.68 & 15.68 & 17.85 \\
${}^{226}\mathrm{Ra}$ & ${}^{14}\mathrm{C}$  & 28.20 & 21.19 & 20.51 & 22.93 \\
${}^{223}\mathrm{Ac}$ & ${}^{14}\mathrm{C}$  & 33.06 & 12.60 & 12.92 & 13.90 \\
${}^{225}\mathrm{Ac}$ & ${}^{14}\mathrm{C}$  & 30.48 & 17.34 & 17.67 & 18.95 \\
${}^{228}\mathrm{Th}$ & ${}^{20}\mathrm{O}$  & 44.72 & 20.72 & 21.19 & 22.95 \\
${}^{230}\mathrm{U}$  & ${}^{22}\mathrm{Ne}$ & 61.39 & 19.57 & 15.40 & 21.24 \\
${}^{231}\mathrm{Pa}$ & ${}^{23}\mathrm{F}$  & 51.88 & 26.00 & 24.71 & 25.17 \\
${}^{230}\mathrm{Th}$ & ${}^{24}\mathrm{Ne}$ & 57.76 & 24.63 & 24.54 & 25.38 \\
${}^{231}\mathrm{Pa}$ & ${}^{24}\mathrm{Ne}$ & 60.41 & 22.89 & 22.94 & 22.50 \\
${}^{232}\mathrm{U}$  & ${}^{24}\mathrm{Ne}$ & 62.31 & 20.39 & 20.27 & 20.93 \\
${}^{233}\mathrm{U}$  & ${}^{24}\mathrm{Ne}$ & 60.49 & 24.82 & 23.82 & 23.71 \\
${}^{234}\mathrm{U}$  & ${}^{24}\mathrm{Ne}$ & 58.83 & 25.07 & 24.99 & 26.35 \\
${}^{235}\mathrm{U}$  & ${}^{24}\mathrm{Ne}$ & 57.36 & 27.62 & 28.16 & 28.78 \\
${}^{233}\mathrm{U}$  & ${}^{25}\mathrm{Ne}$ & 60.70 & 24.82 & 24.81 & 23.98 \\
${}^{234}\mathrm{U}$  & ${}^{26}\mathrm{Ne}$ & 59.41 & 25.30 & 25.45 & 26.65 \\
${}^{234}\mathrm{U}$  & ${}^{28}\mathrm{Mg}$ & 74.11 & 25.53 & 23.85 & 24.99 \\
${}^{236}\mathrm{U}$  & ${}^{28}\mathrm{Mg}$ & 70.73 & 27.58 & 28.14 & 30.01 \\
${}^{238}\mathrm{Pu}$ & ${}^{28}\mathrm{Mg}$ & 75.91 & 25.67 & 23.76 & 25.37 \\
${}^{236}\mathrm{U}$  & ${}^{30}\mathrm{Mg}$ & 72.27 & 27.58 & 28.76 & 28.65 \\
${}^{238}\mathrm{Pu}$ & ${}^{30}\mathrm{Mg}$ & 76.79 & 25.67 & 25.23 & 25.02 \\
${}^{238}\mathrm{Pu}$ & ${}^{32}\mathrm{Si}$ & 91.19 & 25.28 & 25.11 & 23.99 \\
${}^{242}\mathrm{Cm}$ & ${}^{34}\mathrm{Si}$ & 96.54 & 23.24 & 23.59 & 20.89 \\
\hline
\end{tabular}
\end{table}

As shown in Table~\ref{tab:cluster_exp}, DGLM$^{b}$+TabPFN generally provides a substantially improved description of the experimental cluster-radioactivity half-lives compared with DGLM$^{b}$. In particular, the incorporation of TabPFN largely corrects the systematic underestimation exhibited by DGLM$^{b}$ for the experimental dataset. The hybrid-model predictions are also broadly comparable with those obtained from GNL, although noticeable deviations remain for several individual decay channels. To further examine the predictive capability of DGLM$^{b}$+TabPFN for possible cluster emitters, the optimized hybrid model is applied to a set of energetically allowed cluster decay channels for which experimental half-lives are unavailable or have not been compiled in NUBASE2020~\cite{Kondev2021}. The predicted half-lives are compared with those obtained from GNL in Table~\ref{tab:cluster_pre}. The first four columns list the parent nucleus, emitted cluster, decay energy $Q_{\mathrm{c}}$, and orbital angular momentum $\ell$, respectively, while the last two columns give the predicted values of $\log_{10}T_{1/2}$ obtained from DGLM$^{b}$+TabPFN and GNL.

\begin{table}[htbp]
\centering
\caption{Predicted cluster-radioactivity half-lives for possible
cluster emitters.}
\label{tab:cluster_pre}

\small
\setlength{\tabcolsep}{2.5pt}
\renewcommand{\arraystretch}{1.05}

\begin{tabular}{cccccc}
\hline
\multirow{2}{*}{Parent} &
\multirow{2}{*}{Cluster} &
\multirow{2}{*}{$Q_{\mathrm{c}}$} &
\multirow{2}{*}{$\ell$} &
\multicolumn{2}{c}{$\log_{10}T_{1/2}~(\mathrm{s})$} \\
\cline{5-6}
& &  & &
DGLM$^{b}$+TabPFN &
GNL \\
\hline
${}^{219}\mathrm{Rn}$ & ${}^{14}\mathrm{C}$  & 28.10 & 3 & 20.011 & 21.269 \\
${}^{220}\mathrm{Rn}$ & ${}^{14}\mathrm{C}$  & 28.54 & 0 & 17.771 & 20.226 \\
${}^{221}\mathrm{Fr}$ & ${}^{15}\mathrm{N}$  & 34.12 & 3 & 23.074 & 23.250 \\
${}^{223}\mathrm{Ra}$ & ${}^{18}\mathrm{O}$  & 40.30 & 1 & 27.187 & 27.275 \\
${}^{225}\mathrm{Ra}$ & ${}^{14}\mathrm{C}$  & 29.47 & 4 & 18.641 & 20.083 \\
${}^{225}\mathrm{Ra}$ & ${}^{20}\mathrm{O}$  & 40.48 & 1 & 27.931 & 28.779 \\
${}^{226}\mathrm{Ra}$ & ${}^{20}\mathrm{O}$  & 40.82 & 0 & 25.820 & 28.018 \\
${}^{223}\mathrm{Ac}$ & ${}^{15}\mathrm{N}$  & 39.47 & 3 & 15.349 & 15.326 \\
${}^{227}\mathrm{Ac}$ & ${}^{20}\mathrm{O}$  & 43.09 & 1 & 23.935 & 24.779 \\
${}^{229}\mathrm{Ac}$ & ${}^{23}\mathrm{F}$  & 48.35 & 2 & 27.863 & 28.904 \\
${}^{226}\mathrm{Th}$ & ${}^{18}\mathrm{O}$  & 45.73 & 0 & 18.398 & 19.616 \\
${}^{226}\mathrm{Th}$ & ${}^{14}\mathrm{C}$  & 30.55 & 0 & 16.833 & 19.794 \\
${}^{227}\mathrm{Th}$ & ${}^{18}\mathrm{O}$  & 44.20 & 4 & 22.335 & 22.238 \\
${}^{228}\mathrm{Th}$ & ${}^{22}\mathrm{Ne}$ & 55.74 & 0 & 20.480 & 27.471 \\
${}^{229}\mathrm{Th}$ & ${}^{20}\mathrm{O}$  & 43.40 & 2 & 24.397 & 25.396 \\
${}^{229}\mathrm{Th}$ & ${}^{24}\mathrm{Ne}$ & 57.83 & 3 & 25.284 & 25.329 \\
${}^{231}\mathrm{Th}$ & ${}^{24}\mathrm{Ne}$ & 56.25 & 2 & 27.348 & 27.885 \\
${}^{231}\mathrm{Th}$ & ${}^{25}\mathrm{Ne}$ & 56.80 & 2 & 27.834 & 27.608 \\
${}^{232}\mathrm{Th}$ & ${}^{24}\mathrm{Ne}$ & 54.67 & 0 & 28.657 & 30.624 \\
${}^{232}\mathrm{Th}$ & ${}^{26}\mathrm{Ne}$ & 55.91 & 0 & 28.036 & 29.785 \\
${}^{227}\mathrm{Pa}$ & ${}^{18}\mathrm{O}$  & 45.87 & 2 & 20.493 & 20.538 \\
${}^{229}\mathrm{Pa}$ & ${}^{22}\mathrm{Ne}$ & 58.96 & 2 & 18.149 & 23.619 \\
${}^{230}\mathrm{U}$  & ${}^{24}\mathrm{Ne}$ & 61.35 & 0 & 21.307 & 22.548 \\
${}^{232}\mathrm{U}$  & ${}^{28}\mathrm{Mg}$ & 74.32 & 0 & 23.168 & 24.838 \\
${}^{233}\mathrm{U}$  & ${}^{28}\mathrm{Mg}$ & 74.23 & 3 & 24.413 & 24.893 \\
${}^{235}\mathrm{U}$  & ${}^{24}\mathrm{Ne}$ & 57.36 & 1 & 27.581 & 28.776 \\
${}^{235}\mathrm{U}$  & ${}^{25}\mathrm{Ne}$ & 57.68 & 3 & 28.526 & 28.918 \\
${}^{235}\mathrm{U}$  & ${}^{28}\mathrm{Mg}$ & 72.43 & 1 & 26.520 & 27.438 \\
${}^{235}\mathrm{U}$  & ${}^{29}\mathrm{Mg}$ & 72.48 & 3 & 27.791 & 27.889 \\
${}^{236}\mathrm{U}$  & ${}^{24}\mathrm{Ne}$ & 55.95 & 0 & 28.698 & 31.197 \\
${}^{236}\mathrm{U}$  & ${}^{26}\mathrm{Ne}$ & 56.69 & 0 & 28.892 & 31.302 \\
${}^{236}\mathrm{U}$  & ${}^{30}\mathrm{Mg}$ & 72.27 & 0 & 28.104 & 28.650 \\
${}^{238}\mathrm{U}$  & ${}^{30}\mathrm{Mg}$ & 69.46 & 0 & 31.876 & 33.070 \\
${}^{231}\mathrm{Np}$ & ${}^{22}\mathrm{Ne}$ & 61.90 & 3 & 16.387 & 21.801 \\
${}^{233}\mathrm{Np}$ & ${}^{24}\mathrm{Ne}$ & 62.16 & 3 & 22.232 & 22.531 \\
${}^{235}\mathrm{Np}$ & ${}^{28}\mathrm{Mg}$ & 77.10 & 2 & 21.897 & 22.255 \\
${}^{237}\mathrm{Np}$ & ${}^{30}\mathrm{Mg}$ & 74.79 & 2 & 27.192 & 26.403 \\
${}^{237}\mathrm{Pu}$ & ${}^{28}\mathrm{Mg}$ & 77.73 & 1 & 22.196 & 22.834 \\
${}^{237}\mathrm{Pu}$ & ${}^{29}\mathrm{Mg}$ & 77.45 & 3 & 23.790 & 23.691 \\
${}^{237}\mathrm{Pu}$ & ${}^{32}\mathrm{Si}$ & 91.46 & 4 & 25.559 & 23.714 \\
${}^{239}\mathrm{Pu}$ & ${}^{30}\mathrm{Mg}$ & 75.08 & 4 & 28.004 & 27.518 \\
${}^{239}\mathrm{Pu}$ & ${}^{34}\mathrm{Si}$ & 90.87 & 1 & 28.200 & 25.063 \\
${}^{237}\mathrm{Am}$ & ${}^{28}\mathrm{Mg}$ & 79.85 & 2 & 21.000 & 21.494 \\
${}^{239}\mathrm{Am}$ & ${}^{32}\mathrm{Si}$ & 94.50 & 3 & 23.203 & 21.363 \\
${}^{241}\mathrm{Am}$ & ${}^{34}\mathrm{Si}$ & 93.96 & 3 & 25.761 & 22.551 \\
${}^{240}\mathrm{Cm}$ & ${}^{32}\mathrm{Si}$ & 97.55 & 0 & 19.942 & 19.183 \\
${}^{241}\mathrm{Cm}$ & ${}^{32}\mathrm{Si}$ & 95.39 & 4 & 23.400 & 21.816 \\
${}^{243}\mathrm{Cm}$ & ${}^{34}\mathrm{Si}$ & 94.79 & 2 & 25.992 & 23.093 \\
${}^{244}\mathrm{Cm}$ & ${}^{34}\mathrm{Si}$ & 93.17 & 0 & 26.497 & 25.183 \\
\hline
\end{tabular}
\end{table}

As shown in Table~\ref{tab:cluster_pre}, the DGLM$^{b}$+TabPFN and GNL predictions show reasonable agreement for a number of candidate decay channels, whereas relatively large differences remain for several cases. Such discrepancies become more apparent for some heavier emitted clusters, reflecting the increased uncertainty associated with extrapolations beyond the experimentally known dataset. Therefore, these results should be regarded as theoretical predictions for possible cluster emitters and require.

\section{SUMMARY}

In this work, the deformed Gamow-like model (DGLM) is combined with the Tabular Prior-Data Fitted Network (TabPFN) to establish a unified framework for predicting the half-lives of two-proton radioactivity, proton radioactivity, $\alpha$ decay, and cluster radioactivity. A total of 583 decay cases are considered, including 17 two-proton, 42 proton, 498 $\alpha$-decay, and 26 cluster-radioactivity cases. For the complete dataset, the $\sigma_{\mathrm{RMS}}$ values obtained from DGLM$^{a}$, DGLM$^{a}$+TabPFN, DGLM$^{b}$, and DGLM$^{b}$+TabPFN are 2.474, 0.516, 2.383, and 0.423, respectively. DGLM$^{b}$+TabPFN yields the smallest overall $\sigma_{\mathrm{RMS}}$, corresponding to a reduction of approximately $82.2\%$ relative to DGLM$^{b}$.

For the different decay modes, the effective nuclear-radius parameter $r_0$ and the hindrance factor $h$ are optimized by the least-squares method. The optimal $r_0$ values for two-proton radioactivity, proton radioactivity, $\alpha$ decay, and cluster radioactivity are 1.24, 1.28, 1.30, and $1.27~\mathrm{fm}$, respectively. For $\alpha$ decay, the optimized hindrance factors are $h_{\mathrm{EO}}=h_{\mathrm{OE}}=0.220$ and $h_{\mathrm{OO}}=0.510$, while $h=1.18$ is for odd-$A$ cluster cases. Incorporating TabPFN substantially improves the agreement between the calculated and experimental half-lives. In particular, the $\sigma_{\mathrm{RMS}}$ values for proton radioactivity and $\alpha$ decay are reduced by approximately $80.6\%$ and $87.6\%$, respectively. For two-proton radioactivity, DGLM$^{b}$+TabPFN also improves the overall agreement with the experimental data, although relatively large deviations remain for some nuclei. For cluster radioactivity, the hybrid model substantially improves the description of the experimental half-lives compared with DGLM$^{b}$ and provides predictions for possible cluster emitters that can be compared with those obtained from GNL.

For the 498 $\alpha$-decay cases, the $\sigma_{\mathrm{RMS}}$ values for the training set, test set, and entire dataset are 0.208, 0.305, and 0.240, respectively, suggesting reasonable generalization to the test data. In addition, DGLM$^{b}$+TabPFN reproduces the systematic evolution of $\alpha$-decay half-lives along the $Z=84$--87 isotopic chains, including the pronounced variation in the vicinity of the neutron shell closure at $N=126$. Its extrapolation to the superheavy isotopic chains ($Z=117$--120) also exhibits a pronounced change in the predicted half-lives around $N=184$, consistent with the expected influence of the neutron shell closure in this region. However, because experimental half-lives are unavailable for these extremely neutron-rich superheavy nuclei, this behavior should be regarded as a theoretical extrapolation rather than direct evidence for a shell closure.

Overall, the DGLM$^{b}$+TabPFN hybrid framework substantially improves the description of radioactive-decay half-lives while retaining the physically motivated structure of the conventional DGLM. The present results demonstrate its applicability to multiple radioactive-decay modes and its potential for predicting the decay properties of nuclei beyond the experimentally known region.

\begin{center}
    \textbf{ACKNOWLEDGMENTS}
\end{center}

This work is supported by Yunnan Fundamental Research Projects (No. 202501AT070067, 202401AU070074 and 202501CF070189), Yunnan Provincial Xing Dian Talent Support Program (Young Talents Special Program, (Young Talents Special Program, No. XDYC-QNRC-2023-0162), Kunming University Talent Introduction Research Project (No. YJL24019), Yunnan Provincial Department of Education Scientific Research Fund Project (No. 2025Y1055 and 2025Y1042), the Special Basic Cooperative Research Programs of Yunnan ProvincialUndergraduate Universities’ Association (NO. 202101BA070001-144), the Program for Frontier Research Team of Kunming University 2023, National Natural Science Foundation of China (No. 12063006), National College Student Innovation and Entrepreneurship Training Program (No. 202511393011, 202511393012, and 202511393015), Yunnan Province College Student Innovation and Entrepreneurship Training Program (No. S202511393003, S202511393043, and S202511393044), and Xing Dingyu Academician Workstation of Yunnan Province (No. 202605AF350035).


\nocite{*}

\bibliography{apssamp}

\begin{thebibliography}{62}%
\makeatletter
\providecommand \@ifxundefined [1]{%
 \@ifx{#1\undefined}
}%
\providecommand \@ifnum [1]{%
 \ifnum #1\expandafter \@firstoftwo
 \else \expandafter \@secondoftwo
 \fi
}%
\providecommand \@ifx [1]{%
 \ifx #1\expandafter \@firstoftwo
 \else \expandafter \@secondoftwo
 \fi
}%
\providecommand \natexlab [1]{#1}%
\providecommand \enquote  [1]{``#1''}%
\providecommand \bibnamefont  [1]{#1}%
\providecommand \bibfnamefont [1]{#1}%
\providecommand \citenamefont [1]{#1}%
\providecommand \href@noop [0]{\@secondoftwo}%
\providecommand \href [0]{\begingroup \@sanitize@url \@href}%
\providecommand \@href[1]{\@@startlink{#1}\@@href}%
\providecommand \@@href[1]{\endgroup#1\@@endlink}%
\providecommand \@sanitize@url [0]{\catcode `\\12\catcode `\$12\catcode `\&12\catcode `\#12\catcode `\^12\catcode `\_12\catcode `\%12\relax}%
\providecommand \@@startlink[1]{}%
\providecommand \@@endlink[0]{}%
\providecommand \url  [0]{\begingroup\@sanitize@url \@url }%
\providecommand \@url [1]{\endgroup\@href {#1}{\urlprefix }}%
\providecommand \urlprefix  [0]{URL }%
\providecommand \Eprint [0]{\href }%
\providecommand \doibase [0]{https://doi.org/}%
\providecommand \selectlanguage [0]{\@gobble}%
\providecommand \bibinfo  [0]{\@secondoftwo}%
\providecommand \bibfield  [0]{\@secondoftwo}%
\providecommand \translation [1]{[#1]}%
\providecommand \BibitemOpen [0]{}%
\providecommand \bibitemStop [0]{}%
\providecommand \bibitemNoStop [0]{.\EOS\space}%
\providecommand \EOS [0]{\spacefactor3000\relax}%
\providecommand \BibitemShut  [1]{\csname bibitem#1\endcsname}%
\let\auto@bib@innerbib\@empty
\bibitem [{\citenamefont {Hu}\ and\ \citenamefont {Wu}(2026)}]{Jinyu2026}%
  \BibitemOpen
  \bibfield  {author} {\bibinfo {author} {\bibfnamefont {J.}~\bibnamefont {Hu}}\ and\ \bibinfo {author} {\bibfnamefont {C.}~\bibnamefont {Wu}},\ }\bibfield  {title} {\bibinfo {title} {$\alpha$ decay systematics for superheavy nucleus: The effect of deformation of daughter nucleus},\ }\href@noop {} {\bibfield  {journal} {\bibinfo  {journal} {Nuclear Physics A}\ ,\ \bibinfo {pages} {123379}} (\bibinfo {year} {2026})}\BibitemShut {NoStop}%
\bibitem [{\citenamefont {Xiao}\ \emph {et~al.}(2026)\citenamefont {Xiao}, \citenamefont {Qi}, \citenamefont {Yu}, \citenamefont {Yang},\ and\ \citenamefont {Hu}}]{xiao2026}%
  \BibitemOpen
  \bibfield  {author} {\bibinfo {author} {\bibfnamefont {X.}~\bibnamefont {Xiao}}, \bibinfo {author} {\bibfnamefont {P.}~\bibnamefont {Qi}}, \bibinfo {author} {\bibfnamefont {G.}~\bibnamefont {Yu}}, \bibinfo {author} {\bibfnamefont {H.}~\bibnamefont {Yang}},\ and\ \bibinfo {author} {\bibfnamefont {Q.}~\bibnamefont {Hu}},\ }\bibfield  {title} {\bibinfo {title} {Bayesian optimization and nonlocal effects method for $\alpha$ decay of superheavy nuclei based on cppm},\ }\href@noop {} {\bibfield  {journal} {\bibinfo  {journal} {Physica Scripta}\ } (\bibinfo {year} {2026})}\BibitemShut {NoStop}%
\bibitem [{\citenamefont {Qi}\ \emph {et~al.}(2026{\natexlab{a}})\citenamefont {Qi}, \citenamefont {Xiao}, \citenamefont {Yu}, \citenamefont {Yang},\ and\ \citenamefont {Hu}}]{Qi2026}%
  \BibitemOpen
  \bibfield  {author} {\bibinfo {author} {\bibfnamefont {P.}~\bibnamefont {Qi}}, \bibinfo {author} {\bibfnamefont {X.}~\bibnamefont {Xiao}}, \bibinfo {author} {\bibfnamefont {G.}~\bibnamefont {Yu}}, \bibinfo {author} {\bibfnamefont {H.}~\bibnamefont {Yang}},\ and\ \bibinfo {author} {\bibfnamefont {Q.}~\bibnamefont {Hu}},\ }\bibfield  {title} {\bibinfo {title} {Systematic study of the $\alpha$-particle preformation factor in the theory of $\alpha$ decay based on the tabular prior-data fitted network ({TabPFN})},\ }\href@noop {} {\bibfield  {journal} {\bibinfo  {journal} {Physical Review C}\ }\textbf {\bibinfo {volume} {113}},\ \bibinfo {pages} {054606} (\bibinfo {year} {2026}{\natexlab{a}})}\BibitemShut {NoStop}%
\bibitem [{\citenamefont {Yuan}\ \emph {et~al.}(2026{\natexlab{a}})\citenamefont {Yuan}, \citenamefont {Qi}, \citenamefont {Xiao}, \citenamefont {Wang}, \citenamefont {He}, \citenamefont {Long}, \citenamefont {Duan}, \citenamefont {Dai}, \citenamefont {Yan}, \citenamefont {Yu},\ and\ \citenamefont {Yang}}]{Yuan2026}%
  \BibitemOpen
  \bibfield  {author} {\bibinfo {author} {\bibfnamefont {Q.}~\bibnamefont {Yuan}}, \bibinfo {author} {\bibfnamefont {P.}~\bibnamefont {Qi}}, \bibinfo {author} {\bibfnamefont {X.}~\bibnamefont {Xiao}}, \bibinfo {author} {\bibfnamefont {X.}~\bibnamefont {Wang}}, \bibinfo {author} {\bibfnamefont {J.}~\bibnamefont {He}}, \bibinfo {author} {\bibfnamefont {G.}~\bibnamefont {Long}}, \bibinfo {author} {\bibfnamefont {Z.}~\bibnamefont {Duan}}, \bibinfo {author} {\bibfnamefont {Y.}~\bibnamefont {Dai}}, \bibinfo {author} {\bibfnamefont {R.}~\bibnamefont {Yan}}, \bibinfo {author} {\bibfnamefont {G.}~\bibnamefont {Yu}},\ and\ \bibinfo {author} {\bibfnamefont {H.}~\bibnamefont {Yang}},\ }\bibfield  {title} {\bibinfo {title} {Machine learning-driven high-precision model for $\alpha$ decay energy and half-life prediction of superheavy nuclei},\ }\href@noop {} {\bibfield  {journal} {\bibinfo  {journal} {Physica Scripta}\ ,\ \bibinfo {pages} {176004}} (\bibinfo {year} {2026}{\natexlab{a}})}\BibitemShut {NoStop}%
\bibitem [{\citenamefont {Ni}\ \emph {et~al.}(2008)\citenamefont {Ni}, \citenamefont {Ren}, \citenamefont {Dong},\ and\ \citenamefont {Xu}}]{Ni2008}%
  \BibitemOpen
  \bibfield  {author} {\bibinfo {author} {\bibfnamefont {D.}~\bibnamefont {Ni}}, \bibinfo {author} {\bibfnamefont {Z.}~\bibnamefont {Ren}}, \bibinfo {author} {\bibfnamefont {T.}~\bibnamefont {Dong}},\ and\ \bibinfo {author} {\bibfnamefont {C.}~\bibnamefont {Xu}},\ }\bibfield  {title} {\bibinfo {title} {Unified formula of half-lives for $\alpha$ decay and cluster radioactivity},\ }\href@noop {} {\bibfield  {journal} {\bibinfo  {journal} {Physical Review C}\ ,\ \bibinfo {pages} {044310}} (\bibinfo {year} {2008})}\BibitemShut {NoStop}%
\bibitem [{\citenamefont {Qi}\ \emph {et~al.}(2009)\citenamefont {Qi}, \citenamefont {Xu}, \citenamefont {Liotta},\ and\ \citenamefont {Wyss}}]{Qi2009}%
  \BibitemOpen
  \bibfield  {author} {\bibinfo {author} {\bibfnamefont {C.}~\bibnamefont {Qi}}, \bibinfo {author} {\bibfnamefont {F.}~\bibnamefont {Xu}}, \bibinfo {author} {\bibfnamefont {R.}~\bibnamefont {Liotta}},\ and\ \bibinfo {author} {\bibfnamefont {R.}~\bibnamefont {Wyss}},\ }\bibfield  {title} {\bibinfo {title} {Universal decay law in charged-particle emission and exotic cluster radioactivity.},\ }\href@noop {} {\bibfield  {journal} {\bibinfo  {journal} {Physical Review Letters}\ ,\ \bibinfo {pages} {072501(1}} (\bibinfo {year} {2009})}\BibitemShut {NoStop}%
\bibitem [{\citenamefont {Santhosh}\ and\ \citenamefont {Nithya}(2018)}]{Santhosh2018}%
  \BibitemOpen
  \bibfield  {author} {\bibinfo {author} {\bibfnamefont {K.}~\bibnamefont {Santhosh}}\ and\ \bibinfo {author} {\bibfnamefont {C.}~\bibnamefont {Nithya}},\ }\bibfield  {title} {\bibinfo {title} {Systematic studies of $\alpha$ and heavy-cluster emissions from superheavy nuclei},\ }\href@noop {} {\bibfield  {journal} {\bibinfo  {journal} {Physical Review C}\ ,\ \bibinfo {pages} {064616}} (\bibinfo {year} {2018})}\BibitemShut {NoStop}%
\bibitem [{\citenamefont {Teruya}\ \emph {et~al.}(2016)\citenamefont {Teruya}, \citenamefont {Duarte},\ and\ \citenamefont {Rodrigues}}]{N.Teruya2016}%
  \BibitemOpen
  \bibfield  {author} {\bibinfo {author} {\bibfnamefont {N.}~\bibnamefont {Teruya}}, \bibinfo {author} {\bibfnamefont {S.~B.}\ \bibnamefont {Duarte}},\ and\ \bibinfo {author} {\bibfnamefont {M.~M.~N.}\ \bibnamefont {Rodrigues}},\ }\bibfield  {title} {\bibinfo {title} {Nonlocality effect in the tunneling of one-proton radioactivity},\ }\href@noop {} {\bibfield  {journal} {\bibinfo  {journal} {Physical Review C}\ ,\ \bibinfo {pages} {024606}} (\bibinfo {year} {2016})}\BibitemShut {NoStop}%
\bibitem [{\citenamefont {Santhosh}(2022)}]{K.p.2022}%
  \BibitemOpen
  \bibfield  {author} {\bibinfo {author} {\bibfnamefont {K.~P.}\ \bibnamefont {Santhosh}},\ }\bibfield  {title} {\bibinfo {title} {Two-proton radioactivity within a coulomb and proximity potential model for deformed nuclei},\ }\href@noop {} {\bibfield  {journal} {\bibinfo  {journal} {Physical Review C}\ ,\ \bibinfo {pages} {054604}} (\bibinfo {year} {2022})}\BibitemShut {NoStop}%
\bibitem [{\citenamefont {Zdeb}\ \emph {et~al.}(2013{\natexlab{a}})\citenamefont {Zdeb}, \citenamefont {Warda},\ and\ \citenamefont {Pomorski}}]{Zdeb.PRC2013}%
  \BibitemOpen
  \bibfield  {author} {\bibinfo {author} {\bibfnamefont {A.}~\bibnamefont {Zdeb}}, \bibinfo {author} {\bibfnamefont {M.}~\bibnamefont {Warda}},\ and\ \bibinfo {author} {\bibfnamefont {K.}~\bibnamefont {Pomorski}},\ }\bibfield  {title} {\bibinfo {title} {Half-lives for $\alpha$ and cluster radioactivity within a gamow-like model},\ }\href@noop {} {\bibfield  {journal} {\bibinfo  {journal} {Physical Review C}\ }\textbf {\bibinfo {volume} {87}},\ \bibinfo {pages} {024308} (\bibinfo {year} {2013}{\natexlab{a}})}\BibitemShut {NoStop}%
\bibitem [{\citenamefont {Zhao}\ and\ \citenamefont {Bao}(2026)}]{Zhao2026}%
  \BibitemOpen
  \bibfield  {author} {\bibinfo {author} {\bibfnamefont {T.~L.}\ \bibnamefont {Zhao}}\ and\ \bibinfo {author} {\bibfnamefont {X.~J.}\ \bibnamefont {Bao}},\ }\bibfield  {title} {\bibinfo {title} {Exploring the high precision and interpretability of deep learning in the calculation of $\alpha$-decay half-lives},\ }\href {https://doi.org/10.1016/j.physletb.2026.140609} {\bibfield  {journal} {\bibinfo  {journal} {Physics Letters B}\ }\textbf {\bibinfo {volume} {879}},\ \bibinfo {pages} {140609} (\bibinfo {year} {2026})}\BibitemShut {NoStop}%
\bibitem [{\citenamefont {Zhang}\ \emph {et~al.}(2026)\citenamefont {Zhang}, \citenamefont {Li}, \citenamefont {Li}, \citenamefont {Zhong},\ and\ \citenamefont {Yuan}}]{Zhang2026}%
  \BibitemOpen
  \bibfield  {author} {\bibinfo {author} {\bibfnamefont {Y.}~\bibnamefont {Zhang}}, \bibinfo {author} {\bibfnamefont {Z.}~\bibnamefont {Li}}, \bibinfo {author} {\bibfnamefont {K.}~\bibnamefont {Li}}, \bibinfo {author} {\bibfnamefont {J.}~\bibnamefont {Zhong}},\ and\ \bibinfo {author} {\bibfnamefont {C.}~\bibnamefont {Yuan}},\ }\bibfield  {title} {\bibinfo {title} {From $\alpha$ decay to cluster decay: An extreme case of transfer learning},\ }\href@noop {} {\bibfield  {journal} {\bibinfo  {journal} {Physical Review C}\ }\textbf {\bibinfo {volume} {114}},\ \bibinfo {pages} {014314} (\bibinfo {year} {2026})}\BibitemShut {NoStop}%
\bibitem [{\citenamefont {Jain}\ \emph {et~al.}(2026)\citenamefont {Jain}, \citenamefont {Bhuyan}, \citenamefont {Jain},\ and\ \citenamefont {Kumar}}]{Jain2026}%
  \BibitemOpen
  \bibfield  {author} {\bibinfo {author} {\bibfnamefont {N.}~\bibnamefont {Jain}}, \bibinfo {author} {\bibfnamefont {M.}~\bibnamefont {Bhuyan}}, \bibinfo {author} {\bibfnamefont {D.}~\bibnamefont {Jain}},\ and\ \bibinfo {author} {\bibfnamefont {R.}~\bibnamefont {Kumar}},\ }\bibfield  {title} {\bibinfo {title} {Systematics of double-$\alpha$-decay half-lives using machine-learning regression},\ }\href@noop {} {\bibfield  {journal} {\bibinfo  {journal} {Physical Review C}\ }\textbf {\bibinfo {volume} {114}},\ \bibinfo {pages} {014332} (\bibinfo {year} {2026})}\BibitemShut {NoStop}%
\bibitem [{\citenamefont {Qi}\ \emph {et~al.}(2026{\natexlab{b}})\citenamefont {Qi}, \citenamefont {Xiao}, \citenamefont {Yu}, \citenamefont {Yang},\ and\ \citenamefont {Hu}}]{Qi2026EPJP}%
  \BibitemOpen
  \bibfield  {author} {\bibinfo {author} {\bibfnamefont {P.}~\bibnamefont {Qi}}, \bibinfo {author} {\bibfnamefont {X.}~\bibnamefont {Xiao}}, \bibinfo {author} {\bibfnamefont {G.}~\bibnamefont {Yu}}, \bibinfo {author} {\bibfnamefont {H.}~\bibnamefont {Yang}},\ and\ \bibinfo {author} {\bibfnamefont {Q.}~\bibnamefont {Hu}},\ }\bibfield  {title} {\bibinfo {title} {Systematic study of $\alpha$-decay half-lives of superheavy nuclei based on coulomb and proximity potential models with temperature effects},\ }\href {https://doi.org/10.1140/epjp/s13360-026-07528-5} {\bibfield  {journal} {\bibinfo  {journal} {The European Physical Journal Plus}\ }\textbf {\bibinfo {volume} {141}},\ \bibinfo {pages} {341} (\bibinfo {year} {2026}{\natexlab{b}})}\BibitemShut {NoStop}%
\bibitem [{\citenamefont {Wu}(2025)}]{Jinyu2025}%
  \BibitemOpen
  \bibfield  {author} {\bibinfo {author} {\bibfnamefont {J.~H. .~C.}\ \bibnamefont {Wu}},\ }\bibfield  {title} {\bibinfo {title} {Nonlocality effect in $\alpha$ decay half-lives for even-even nuclei within a two potential approach},\ }\href@noop {} {\bibfield  {journal} {\bibinfo  {journal} {The European Physical Journal A}\ ,\ \bibinfo {pages} {1}} (\bibinfo {year} {2025})}\BibitemShut {NoStop}%
\bibitem [{\citenamefont {Wu}(2026)}]{Jinyu2026_3}%
  \BibitemOpen
  \bibfield  {author} {\bibinfo {author} {\bibfnamefont {J.~H. .~C.}\ \bibnamefont {Wu}},\ }\bibfield  {title} {\bibinfo {title} {Nonlocality effect in the tunneling of $\alpha$ radioactivity with the aid of machine learning},\ }\href@noop {} {\bibfield  {journal} {\bibinfo  {journal} {The European Physical Journal A}\ } (\bibinfo {year} {2026})}\BibitemShut {NoStop}%
\bibitem [{\citenamefont {Zhang}\ and\ \citenamefont {Royer}(2007)}]{Zhang2007}%
  \BibitemOpen
  \bibfield  {author} {\bibinfo {author} {\bibfnamefont {H.~F.}\ \bibnamefont {Zhang}}\ and\ \bibinfo {author} {\bibfnamefont {G.}~\bibnamefont {Royer}},\ }\bibfield  {title} {\bibinfo {title} {Theoretical and experimental $\alpha$ decay half-lives of the heaviest odd-$z$ elements and general predictions},\ }\href@noop {} {\bibfield  {journal} {\bibinfo  {journal} {Physical Review C}\ }\textbf {\bibinfo {volume} {76}},\ \bibinfo {pages} {047304} (\bibinfo {year} {2007})}\BibitemShut {NoStop}%
\bibitem [{\citenamefont {Wan}\ and\ \citenamefont {Fan}(2021)}]{Wan2021}%
  \BibitemOpen
  \bibfield  {author} {\bibinfo {author} {\bibfnamefont {N.}~\bibnamefont {Wan}}\ and\ \bibinfo {author} {\bibfnamefont {J.}~\bibnamefont {Fan}},\ }\bibfield  {title} {\bibinfo {title} {Systematical calculations on $\alpha$-cluster preformation factors and decay half-lives of light nuclei near the recently observed $\alpha$ emitters $^{108}\mathrm{Xe}$ and $^{104}\mathrm{Te}$},\ }\href@noop {} {\bibfield  {journal} {\bibinfo  {journal} {Physical Review C}\ }\textbf {\bibinfo {volume} {104}},\ \bibinfo {pages} {064320} (\bibinfo {year} {2021})}\BibitemShut {NoStop}%
\bibitem [{\citenamefont {Zhang}\ \emph {et~al.}(2006)\citenamefont {Zhang}, \citenamefont {Zuo}, \citenamefont {Li},\ and\ \citenamefont {Royer}}]{Zhang2006}%
  \BibitemOpen
  \bibfield  {author} {\bibinfo {author} {\bibfnamefont {H.}~\bibnamefont {Zhang}}, \bibinfo {author} {\bibfnamefont {W.}~\bibnamefont {Zuo}}, \bibinfo {author} {\bibfnamefont {J.}~\bibnamefont {Li}},\ and\ \bibinfo {author} {\bibfnamefont {G.}~\bibnamefont {Royer}},\ }\bibfield  {title} {\bibinfo {title} {$\alpha$ decay half-lives of new superheavy nuclei within a generalized liquid drop model},\ }\href@noop {} {\bibfield  {journal} {\bibinfo  {journal} {Physical Review C}\ }\textbf {\bibinfo {volume} {74}},\ \bibinfo {pages} {017304} (\bibinfo {year} {2006})}\BibitemShut {NoStop}%
\bibitem [{\citenamefont {Wang}\ \emph {et~al.}(2013)\citenamefont {Wang}, \citenamefont {Zhang},\ and\ \citenamefont {Li}}]{Wang2013}%
  \BibitemOpen
  \bibfield  {author} {\bibinfo {author} {\bibfnamefont {J.~M.}\ \bibnamefont {Wang}}, \bibinfo {author} {\bibfnamefont {H.~F.}\ \bibnamefont {Zhang}},\ and\ \bibinfo {author} {\bibfnamefont {J.~Q.}\ \bibnamefont {Li}},\ }\bibfield  {title} {\bibinfo {title} {$\alpha$ decay half-lives for z = 108, 114, 120, 126 isotopes and n = 162, 184 isotones},\ }\href {https://doi.org/10.1088/0954-3899/40/4/045103} {\bibfield  {journal} {\bibinfo  {journal} {Journal of Physics G: Nuclear and Particle Physics}\ }\textbf {\bibinfo {volume} {40}},\ \bibinfo {pages} {045103} (\bibinfo {year} {2013})}\BibitemShut {NoStop}%
\bibitem [{\citenamefont {Warda}\ \emph {et~al.}(2018)\citenamefont {Warda}, \citenamefont {Zdeb},\ and\ \citenamefont {Robledo}}]{Warda2018}%
  \BibitemOpen
  \bibfield  {author} {\bibinfo {author} {\bibfnamefont {M.}~\bibnamefont {Warda}}, \bibinfo {author} {\bibfnamefont {A.}~\bibnamefont {Zdeb}},\ and\ \bibinfo {author} {\bibfnamefont {L.~M.}\ \bibnamefont {Robledo}},\ }\bibfield  {title} {\bibinfo {title} {Cluster radioactivity in superheavy nuclei},\ }\href {https://doi.org/10.1103/PhysRevC.98.041602} {\bibfield  {journal} {\bibinfo  {journal} {Physical Review C}\ }\textbf {\bibinfo {volume} {98}},\ \bibinfo {pages} {041602(R)} (\bibinfo {year} {2018})}\BibitemShut {NoStop}%
\bibitem [{\citenamefont {Jain}\ \emph {et~al.}(2025)\citenamefont {Jain}, \citenamefont {Jain},\ and\ \citenamefont {Kumar}}]{Jain2025}%
  \BibitemOpen
  \bibfield  {author} {\bibinfo {author} {\bibfnamefont {D.}~\bibnamefont {Jain}}, \bibinfo {author} {\bibfnamefont {N.}~\bibnamefont {Jain}},\ and\ \bibinfo {author} {\bibfnamefont {R.}~\bibnamefont {Kumar}},\ }\bibfield  {title} {\bibinfo {title} {Exploring cluster radioactivity and decay half-lives using different nuclear densities and approaches},\ }\href@noop {} {\bibfield  {journal} {\bibinfo  {journal} {Physical Review C}\ }\textbf {\bibinfo {volume} {111}},\ \bibinfo {pages} {064603} (\bibinfo {year} {2025})}\BibitemShut {NoStop}%
\bibitem [{\citenamefont {Majekodunmi}\ \emph {et~al.}(2022)\citenamefont {Majekodunmi}, \citenamefont {Bhuyan}, \citenamefont {Jain}, \citenamefont {Anwar}, \citenamefont {Abdullah},\ and\ \citenamefont {Kumar}}]{Majekodunmi2022}%
  \BibitemOpen
  \bibfield  {author} {\bibinfo {author} {\bibfnamefont {J.~T.}\ \bibnamefont {Majekodunmi}}, \bibinfo {author} {\bibfnamefont {M.}~\bibnamefont {Bhuyan}}, \bibinfo {author} {\bibfnamefont {D.}~\bibnamefont {Jain}}, \bibinfo {author} {\bibfnamefont {K.}~\bibnamefont {Anwar}}, \bibinfo {author} {\bibfnamefont {N.}~\bibnamefont {Abdullah}},\ and\ \bibinfo {author} {\bibfnamefont {R.}~\bibnamefont {Kumar}},\ }\bibfield  {title} {\bibinfo {title} {Cluster decay half-lives of $^{112--122}\mathrm{Ba}$ isotopes from the ground state and intrinsic excited state using the relativistic mean-field formalism within the preformed-cluster-decay model},\ }\href@noop {} {\bibfield  {journal} {\bibinfo  {journal} {Physical Review C}\ }\textbf {\bibinfo {volume} {105}},\ \bibinfo {pages} {044617} (\bibinfo {year} {2022})}\BibitemShut {NoStop}%
\bibitem [{\citenamefont {Gamow}(1928)}]{Gamow1928}%
  \BibitemOpen
  \bibfield  {author} {\bibinfo {author} {\bibfnamefont {G.}~\bibnamefont {Gamow}},\ }\bibfield  {title} {\bibinfo {title} {Zur quantentheorie des atomkernes},\ }\href@noop {} {\bibfield  {journal} {\bibinfo  {journal} {Zeitschrift für Physik}\ ,\ \bibinfo {pages} {204}} (\bibinfo {year} {1928})}\BibitemShut {NoStop}%
\bibitem [{\citenamefont {Gurney}\ and\ \citenamefont {Condon}(1928)}]{Gurney1928}%
  \BibitemOpen
  \bibfield  {author} {\bibinfo {author} {\bibfnamefont {R.~W.}\ \bibnamefont {Gurney}}\ and\ \bibinfo {author} {\bibfnamefont {E.~U.}\ \bibnamefont {Condon}},\ }\bibfield  {title} {\bibinfo {title} {Wave mechanics and radioactive disintegration},\ }\href {https://doi.org/10.1038/122439a0} {\bibfield  {journal} {\bibinfo  {journal} {Nature}\ }\textbf {\bibinfo {volume} {122}},\ \bibinfo {pages} {439} (\bibinfo {year} {1928})}\BibitemShut {NoStop}%
\bibitem [{\citenamefont {Grigorenko}\ \emph {et~al.}(2000)\citenamefont {Grigorenko}, \citenamefont {Johnson}, \citenamefont {Mukha}, \citenamefont {Thompson},\ and\ \citenamefont {Zhukov}}]{Grigorenko2000}%
  \BibitemOpen
  \bibfield  {author} {\bibinfo {author} {\bibfnamefont {L.~V.}\ \bibnamefont {Grigorenko}}, \bibinfo {author} {\bibfnamefont {R.~C.}\ \bibnamefont {Johnson}}, \bibinfo {author} {\bibfnamefont {I.~G.}\ \bibnamefont {Mukha}}, \bibinfo {author} {\bibfnamefont {I.~J.}\ \bibnamefont {Thompson}},\ and\ \bibinfo {author} {\bibfnamefont {M.~V.}\ \bibnamefont {Zhukov}},\ }\bibfield  {title} {\bibinfo {title} {Theory of two-proton radioactivity with application to ${}^{19}\mathrm{Mg}$ and ${}^{48}\mathrm{Ni}$},\ }\href@noop {} {\bibfield  {journal} {\bibinfo  {journal} {Physical Review Letters}\ }\textbf {\bibinfo {volume} {85}},\ \bibinfo {pages} {22} (\bibinfo {year} {2000})}\BibitemShut {NoStop}%
\bibitem [{\citenamefont {Grigorenko}\ \emph {et~al.}(2001)\citenamefont {Grigorenko}, \citenamefont {Johnson}, \citenamefont {Mukha}, \citenamefont {Thompson},\ and\ \citenamefont {Zhukov}}]{Grigorenko2001}%
  \BibitemOpen
  \bibfield  {author} {\bibinfo {author} {\bibfnamefont {L.~V.}\ \bibnamefont {Grigorenko}}, \bibinfo {author} {\bibfnamefont {R.~C.}\ \bibnamefont {Johnson}}, \bibinfo {author} {\bibfnamefont {I.~G.}\ \bibnamefont {Mukha}}, \bibinfo {author} {\bibfnamefont {I.~J.}\ \bibnamefont {Thompson}},\ and\ \bibinfo {author} {\bibfnamefont {M.~V.}\ \bibnamefont {Zhukov}},\ }\bibfield  {title} {\bibinfo {title} {Two-proton radioactivity and three-body decay: General problems and theoretical approach},\ }\href@noop {} {\bibfield  {journal} {\bibinfo  {journal} {Physical Review C}\ }\textbf {\bibinfo {volume} {64}},\ \bibinfo {pages} {054002} (\bibinfo {year} {2001})}\BibitemShut {NoStop}%
\bibitem [{\citenamefont {Grigorenko}\ and\ \citenamefont {Zhukov}(2003)}]{Grigorenko2003}%
  \BibitemOpen
  \bibfield  {author} {\bibinfo {author} {\bibfnamefont {L.~V.}\ \bibnamefont {Grigorenko}}\ and\ \bibinfo {author} {\bibfnamefont {M.~V.}\ \bibnamefont {Zhukov}},\ }\bibfield  {title} {\bibinfo {title} {Two-proton radioactivity and three-body decay. ii. exploratory studies of lifetimes and correlations},\ }\href@noop {} {\bibfield  {journal} {\bibinfo  {journal} {Physical Review C}\ }\textbf {\bibinfo {volume} {68}},\ \bibinfo {pages} {054005} (\bibinfo {year} {2003})}\BibitemShut {NoStop}%
\bibitem [{\citenamefont {Grigorenko}\ and\ \citenamefont {Zhukov}(2007)}]{Grigorenko2007}%
  \BibitemOpen
  \bibfield  {author} {\bibinfo {author} {\bibfnamefont {L.~V.}\ \bibnamefont {Grigorenko}}\ and\ \bibinfo {author} {\bibfnamefont {M.~V.}\ \bibnamefont {Zhukov}},\ }\bibfield  {title} {\bibinfo {title} {Two-proton radioactivity and three-body decay. iii. integral formulas for decay widths in a simplified semianalytical approach},\ }\href@noop {} {\bibfield  {journal} {\bibinfo  {journal} {Physical Review C}\ }\textbf {\bibinfo {volume} {76}},\ \bibinfo {pages} {014008} (\bibinfo {year} {2007})}\BibitemShut {NoStop}%
\bibitem [{\citenamefont {Ren}\ \emph {et~al.}(2004)\citenamefont {Ren}, \citenamefont {Xu},\ and\ \citenamefont {Wang}}]{Ren2004}%
  \BibitemOpen
  \bibfield  {author} {\bibinfo {author} {\bibfnamefont {Z.}~\bibnamefont {Ren}}, \bibinfo {author} {\bibfnamefont {C.}~\bibnamefont {Xu}},\ and\ \bibinfo {author} {\bibfnamefont {Z.}~\bibnamefont {Wang}},\ }\bibfield  {title} {\bibinfo {title} {New perspective on complex cluster radioactivity of heavy nuclei},\ }\href@noop {} {\bibfield  {journal} {\bibinfo  {journal} {Physical Review C}\ }\textbf {\bibinfo {volume} {70}},\ \bibinfo {pages} {034304} (\bibinfo {year} {2004})}\BibitemShut {NoStop}%
\bibitem [{\citenamefont {Zdeb}\ \emph {et~al.}(2013{\natexlab{b}})\citenamefont {Zdeb}, \citenamefont {Warda},\ and\ \citenamefont {Pomorski}}]{Zdeb2013}%
  \BibitemOpen
  \bibfield  {author} {\bibinfo {author} {\bibfnamefont {A.}~\bibnamefont {Zdeb}}, \bibinfo {author} {\bibfnamefont {M.}~\bibnamefont {Warda}},\ and\ \bibinfo {author} {\bibfnamefont {K.}~\bibnamefont {Pomorski}},\ }\bibfield  {title} {\bibinfo {title} {Half-lives for $\alpha$ and cluster radioactivity in a simple model},\ }\href@noop {} {\bibfield  {journal} {\bibinfo  {journal} {Physica Scripta}\ }\textbf {\bibinfo {volume} {2013}},\ \bibinfo {pages} {014029} (\bibinfo {year} {2013}{\natexlab{b}})}\BibitemShut {NoStop}%
\bibitem [{\citenamefont {Xing}\ \emph {et~al.}(2022)\citenamefont {Xing}, \citenamefont {Qi}, \citenamefont {Cui}, \citenamefont {Gao}, \citenamefont {Wang}, \citenamefont {Gu},\ and\ \citenamefont {Yong}}]{Xing2022IMGL}%
  \BibitemOpen
  \bibfield  {author} {\bibinfo {author} {\bibfnamefont {F.}~\bibnamefont {Xing}}, \bibinfo {author} {\bibfnamefont {H.}~\bibnamefont {Qi}}, \bibinfo {author} {\bibfnamefont {J.}~\bibnamefont {Cui}}, \bibinfo {author} {\bibfnamefont {Y.}~\bibnamefont {Gao}}, \bibinfo {author} {\bibfnamefont {Y.}~\bibnamefont {Wang}}, \bibinfo {author} {\bibfnamefont {J.}~\bibnamefont {Gu}},\ and\ \bibinfo {author} {\bibfnamefont {G.}~\bibnamefont {Yong}},\ }\bibfield  {title} {\bibinfo {title} {An improved gamow-like formula for $\alpha$-decay half-lives},\ }\href@noop {} {\bibfield  {journal} {\bibinfo  {journal} {Nuclear Physics A}\ }\textbf {\bibinfo {volume} {1028}},\ \bibinfo {pages} {122528} (\bibinfo {year} {2022})}\BibitemShut {NoStop}%
\bibitem [{\citenamefont {Azeez}\ \emph {et~al.}(2022)\citenamefont {Azeez}, \citenamefont {Yahya},\ and\ \citenamefont {Saeed}}]{Azeez2022}%
  \BibitemOpen
  \bibfield  {author} {\bibinfo {author} {\bibfnamefont {O.~K.}\ \bibnamefont {Azeez}}, \bibinfo {author} {\bibfnamefont {W.~A.}\ \bibnamefont {Yahya}},\ and\ \bibinfo {author} {\bibfnamefont {A.~A.}\ \bibnamefont {Saeed}},\ }\bibfield  {title} {\bibinfo {title} {Predictions of the $\alpha$-decay half-lives of even--even superheavy nuclei using a modified gamow-like model},\ }\href@noop {} {\bibfield  {journal} {\bibinfo  {journal} {Physica Scripta}\ }\textbf {\bibinfo {volume} {97}},\ \bibinfo {pages} {055302} (\bibinfo {year} {2022})}\BibitemShut {NoStop}%
\bibitem [{\citenamefont {Dong}\ \emph {et~al.}(2009)\citenamefont {Dong}, \citenamefont {Zhang},\ and\ \citenamefont {Royer}}]{Dong2009}%
  \BibitemOpen
  \bibfield  {author} {\bibinfo {author} {\bibfnamefont {J.~M.}\ \bibnamefont {Dong}}, \bibinfo {author} {\bibfnamefont {H.~F.}\ \bibnamefont {Zhang}},\ and\ \bibinfo {author} {\bibfnamefont {G.}~\bibnamefont {Royer}},\ }\bibfield  {title} {\bibinfo {title} {Proton radioactivity within $\alpha$ generalized liquid drop model},\ }\href {https://doi.org/10.1103/PhysRevC.79.054330} {\bibfield  {journal} {\bibinfo  {journal} {Physical Review C}\ }\textbf {\bibinfo {volume} {79}},\ \bibinfo {pages} {054330} (\bibinfo {year} {2009})}\BibitemShut {NoStop}%
\bibitem [{\citenamefont {Hollmann}\ \emph {et~al.}(2025)\citenamefont {Hollmann}, \citenamefont {M{\"u}ller}, \citenamefont {Purucker}, \citenamefont {Krishnakumar}, \citenamefont {K{\"o}rfer}, \citenamefont {Hoo}, \citenamefont {Schirrmeister},\ and\ \citenamefont {Hutter}}]{Noah2025}%
  \BibitemOpen
  \bibfield  {author} {\bibinfo {author} {\bibfnamefont {N.}~\bibnamefont {Hollmann}}, \bibinfo {author} {\bibfnamefont {S.}~\bibnamefont {M{\"u}ller}}, \bibinfo {author} {\bibfnamefont {L.}~\bibnamefont {Purucker}}, \bibinfo {author} {\bibfnamefont {A.}~\bibnamefont {Krishnakumar}}, \bibinfo {author} {\bibfnamefont {M.}~\bibnamefont {K{\"o}rfer}}, \bibinfo {author} {\bibfnamefont {S.~B.}\ \bibnamefont {Hoo}}, \bibinfo {author} {\bibfnamefont {R.~T.}\ \bibnamefont {Schirrmeister}},\ and\ \bibinfo {author} {\bibfnamefont {F.}~\bibnamefont {Hutter}},\ }\bibfield  {title} {\bibinfo {title} {Accurate predictions on small data with a tabular foundation model},\ }\href@noop {} {\bibfield  {journal} {\bibinfo  {journal} {Nature}\ }\textbf {\bibinfo {volume} {637}},\ \bibinfo {pages} {319} (\bibinfo {year} {2025})}\BibitemShut {NoStop}%
\bibitem [{\citenamefont {Liu}\ and\ \citenamefont {Ye}(2025)}]{Si_Yang2025}%
  \BibitemOpen
  \bibfield  {author} {\bibinfo {author} {\bibfnamefont {S.-Y.}\ \bibnamefont {Liu}}\ and\ \bibinfo {author} {\bibfnamefont {H.-J.}\ \bibnamefont {Ye}},\ }\href {https://arxiv.org/abs/2502.02527} {\bibinfo {title} {Tabpfn unleashed: A scalable and effective solution to tabular classification problems}} (\bibinfo {year} {2025}),\ \Eprint {https://arxiv.org/abs/2502.02527} {arXiv:2502.02527 [cs.LG]} \BibitemShut {NoStop}%
\bibitem [{\citenamefont {Yuan}\ \emph {et~al.}(2026{\natexlab{b}})\citenamefont {Yuan}, \citenamefont {Xiao}, \citenamefont {Qi}, \citenamefont {Yang}, \citenamefont {Yu}, \citenamefont {Yang}, \citenamefont {Li},\ and\ \citenamefont {Cai}}]{Yuan2026_2}%
  \BibitemOpen
  \bibfield  {author} {\bibinfo {author} {\bibfnamefont {Q.}~\bibnamefont {Yuan}}, \bibinfo {author} {\bibfnamefont {X.}~\bibnamefont {Xiao}}, \bibinfo {author} {\bibfnamefont {P.}~\bibnamefont {Qi}}, \bibinfo {author} {\bibfnamefont {A.}~\bibnamefont {Yang}}, \bibinfo {author} {\bibfnamefont {G.}~\bibnamefont {Yu}}, \bibinfo {author} {\bibfnamefont {H.}~\bibnamefont {Yang}}, \bibinfo {author} {\bibfnamefont {Z.}~\bibnamefont {Li}},\ and\ \bibinfo {author} {\bibfnamefont {Y.}~\bibnamefont {Cai}},\ }\bibfield  {title} {\bibinfo {title} {Physics-guided residual correction of $\ensuremath{\alpha}$-decay half-lives based on the effective liquid drop model},\ }\href {https://doi.org/10.1103/7gm3-2y7f} {\bibfield  {journal} {\bibinfo  {journal} {Phys. Rev. C}\ } (\bibinfo {year} {2026}{\natexlab{b}})}\BibitemShut {NoStop}%
\bibitem [{\citenamefont {Anyas-Weiss}\ \emph {et~al.}(1974)\citenamefont {Anyas-Weiss}, \citenamefont {Cornell}, \citenamefont {Fisher}, \citenamefont {Hudson}, \citenamefont {Menchaca-Rocha}, \citenamefont {Panagiotou}, \citenamefont {Scott}, \citenamefont {Strottman}, \citenamefont {Brink}, \citenamefont {Buck}, \citenamefont {Ellis},\ and\ \citenamefont {Engeland}}]{Anyas-Weiss1974}%
  \BibitemOpen
  \bibfield  {author} {\bibinfo {author} {\bibfnamefont {N.}~\bibnamefont {Anyas-Weiss}}, \bibinfo {author} {\bibfnamefont {J.~C.}\ \bibnamefont {Cornell}}, \bibinfo {author} {\bibfnamefont {P.~S.}\ \bibnamefont {Fisher}}, \bibinfo {author} {\bibfnamefont {P.~N.}\ \bibnamefont {Hudson}}, \bibinfo {author} {\bibfnamefont {A.}~\bibnamefont {Menchaca-Rocha}}, \bibinfo {author} {\bibfnamefont {A.~D.}\ \bibnamefont {Panagiotou}}, \bibinfo {author} {\bibfnamefont {D.~K.}\ \bibnamefont {Scott}}, \bibinfo {author} {\bibfnamefont {D.}~\bibnamefont {Strottman}}, \bibinfo {author} {\bibfnamefont {D.~M.}\ \bibnamefont {Brink}}, \bibinfo {author} {\bibfnamefont {B.}~\bibnamefont {Buck}}, \bibinfo {author} {\bibfnamefont {P.~J.}\ \bibnamefont {Ellis}},\ and\ \bibinfo {author} {\bibfnamefont {T.}~\bibnamefont {Engeland}},\ }\bibfield  {title} {\bibinfo {title} {Nuclear structure of light nuclei using the selectivity of high energy transfer reactions with heavy ions},\ }\href@noop {} {\bibfield  {journal} {\bibinfo
  {journal} {Physics Reports}\ }\textbf {\bibinfo {volume} {12}},\ \bibinfo {pages} {201} (\bibinfo {year} {1974})}\BibitemShut {NoStop}%
\bibitem [{\citenamefont {Xu}\ and\ \citenamefont {Ren}(2006)}]{Xu2006}%
  \BibitemOpen
  \bibfield  {author} {\bibinfo {author} {\bibfnamefont {C.}~\bibnamefont {Xu}}\ and\ \bibinfo {author} {\bibfnamefont {Z.}~\bibnamefont {Ren}},\ }\bibfield  {title} {\bibinfo {title} {New deformed model of $\alpha$-decay half-lives with a microscopic potential},\ }\href@noop {} {\bibfield  {journal} {\bibinfo  {journal} {Physical Review C}\ }\textbf {\bibinfo {volume} {73}},\ \bibinfo {pages} {041301(R)} (\bibinfo {year} {2006})}\BibitemShut {NoStop}%
\bibitem [{\citenamefont {M{\"o}ller}\ \emph {et~al.}(2016)\citenamefont {M{\"o}ller}, \citenamefont {Sierk}, \citenamefont {Ichikawa},\ and\ \citenamefont {Sagawa}}]{P2016}%
  \BibitemOpen
  \bibfield  {author} {\bibinfo {author} {\bibfnamefont {P.}~\bibnamefont {M{\"o}ller}}, \bibinfo {author} {\bibfnamefont {A.~J.}\ \bibnamefont {Sierk}}, \bibinfo {author} {\bibfnamefont {T.}~\bibnamefont {Ichikawa}},\ and\ \bibinfo {author} {\bibfnamefont {H.}~\bibnamefont {Sagawa}},\ }\bibfield  {title} {\bibinfo {title} {Nuclear ground-state masses and deformations: {FRDM}(2012)},\ }\href@noop {} {\bibfield  {journal} {\bibinfo  {journal} {Atomic Data and Nuclear Data Tables}\ }\textbf {\bibinfo {volume} {109--110}},\ \bibinfo {pages} {1} (\bibinfo {year} {2016})}\BibitemShut {NoStop}%
\bibitem [{\citenamefont {Blendowske}\ and\ \citenamefont {Walliser}(1988)}]{Blendowske1988}%
  \BibitemOpen
  \bibfield  {author} {\bibinfo {author} {\bibfnamefont {R.}~\bibnamefont {Blendowske}}\ and\ \bibinfo {author} {\bibfnamefont {H.}~\bibnamefont {Walliser}},\ }\bibfield  {title} {\bibinfo {title} {Systematics of cluster-radioactivity-decay constants as suggested by microscopic calculations},\ }\href@noop {} {\bibfield  {journal} {\bibinfo  {journal} {Physical Review Letters}\ }\textbf {\bibinfo {volume} {61}},\ \bibinfo {pages} {1930} (\bibinfo {year} {1988})}\BibitemShut {NoStop}%
\bibitem [{\citenamefont {Lee}(2026)}]{Kyungeun2026}%
  \BibitemOpen
  \bibfield  {author} {\bibinfo {author} {\bibfnamefont {K.}~\bibnamefont {Lee}},\ }\bibfield  {title} {\bibinfo {title} {Multitabpfn: Codebook-based extensions of tabpfn for high-class-count tabular classification},\ }\href@noop {} {\bibfield  {journal} {\bibinfo  {journal} {Neural networks : the official journal of the International Neural Network Society}\ ,\ \bibinfo {pages} {108932}} (\bibinfo {year} {2026})}\BibitemShut {NoStop}%
\bibitem [{\citenamefont {Chen}\ \emph {et~al.}(2019)\citenamefont {Chen}, \citenamefont {Xu}, \citenamefont {Deng}, \citenamefont {Li}, \citenamefont {He},\ and\ \citenamefont {Chu}}]{Chen2019}%
  \BibitemOpen
  \bibfield  {author} {\bibinfo {author} {\bibfnamefont {J.-L.}\ \bibnamefont {Chen}}, \bibinfo {author} {\bibfnamefont {J.-Y.}\ \bibnamefont {Xu}}, \bibinfo {author} {\bibfnamefont {J.-G.}\ \bibnamefont {Deng}}, \bibinfo {author} {\bibfnamefont {X.-H.}\ \bibnamefont {Li}}, \bibinfo {author} {\bibfnamefont {B.}~\bibnamefont {He}},\ and\ \bibinfo {author} {\bibfnamefont {P.-C.}\ \bibnamefont {Chu}},\ }\bibfield  {title} {\bibinfo {title} {New geiger--nuttall law for proton radioactivity},\ }\href {https://doi.org/10.1140/epja/i2019-12927-7} {\bibfield  {journal} {\bibinfo  {journal} {The European Physical Journal A}\ }\textbf {\bibinfo {volume} {55}},\ \bibinfo {pages} {214} (\bibinfo {year} {2019})}\BibitemShut {NoStop}%
\bibitem [{\citenamefont {Liu}\ \emph {et~al.}(2021)\citenamefont {Liu}, \citenamefont {Zou}, \citenamefont {Pan}, \citenamefont {He},\ and\ \citenamefont {Li}}]{Liu2021}%
  \BibitemOpen
  \bibfield  {author} {\bibinfo {author} {\bibfnamefont {H.-M.}\ \bibnamefont {Liu}}, \bibinfo {author} {\bibfnamefont {Y.-T.}\ \bibnamefont {Zou}}, \bibinfo {author} {\bibfnamefont {J.-L.}\ \bibnamefont {Pan}, \bibfnamefont {Xiao~andChen}}, \bibinfo {author} {\bibfnamefont {B.}~\bibnamefont {He}},\ and\ \bibinfo {author} {\bibfnamefont {X.-H.}\ \bibnamefont {Li}},\ }\bibfield  {title} {\bibinfo {title} {New geiger--nuttall law for two-proton radioactivity},\ }\href {https://doi.org/10.1088/1674-1137/abd01e} {\bibfield  {journal} {\bibinfo  {journal} {Chinese Physics C}\ }\textbf {\bibinfo {volume} {45}},\ \bibinfo {pages} {024108} (\bibinfo {year} {2021})}\BibitemShut {NoStop}%
\bibitem [{\citenamefont {Zdeb}\ \emph {et~al.}(2016)\citenamefont {Zdeb}, \citenamefont {Warda}, \citenamefont {Petrache},\ and\ \citenamefont {Pomorski}}]{Zdeb2016}%
  \BibitemOpen
  \bibfield  {author} {\bibinfo {author} {\bibfnamefont {A.}~\bibnamefont {Zdeb}}, \bibinfo {author} {\bibfnamefont {M.}~\bibnamefont {Warda}}, \bibinfo {author} {\bibfnamefont {C.~M.}\ \bibnamefont {Petrache}},\ and\ \bibinfo {author} {\bibfnamefont {K.}~\bibnamefont {Pomorski}},\ }\bibfield  {title} {\bibinfo {title} {Proton emission half-lives within a gamow-like model},\ }\href@noop {} {\bibfield  {journal} {\bibinfo  {journal} {The European Physical Journal A}\ }\textbf {\bibinfo {volume} {52}},\ \bibinfo {pages} {323} (\bibinfo {year} {2016})}\BibitemShut {NoStop}%
\bibitem [{\citenamefont {Grigorenko}(2009)}]{Grigorenko2009}%
  \BibitemOpen
  \bibfield  {author} {\bibinfo {author} {\bibfnamefont {L.~V.}\ \bibnamefont {Grigorenko}},\ }\bibfield  {title} {\bibinfo {title} {Theoretical study of two-proton radioactivity. status, predictions, and applications},\ }\href@noop {} {\bibfield  {journal} {\bibinfo  {journal} {Physics of Particles and Nuclei}\ }\textbf {\bibinfo {volume} {40}},\ \bibinfo {pages} {674} (\bibinfo {year} {2009})}\BibitemShut {NoStop}%
\bibitem [{\citenamefont {Rotureau}\ \emph {et~al.}(2006)\citenamefont {Rotureau}, \citenamefont {Oko{\l}owicz},\ and\ \citenamefont {P{\l}oszajczak}}]{Rotureau2006}%
  \BibitemOpen
  \bibfield  {author} {\bibinfo {author} {\bibfnamefont {J.}~\bibnamefont {Rotureau}}, \bibinfo {author} {\bibfnamefont {J.}~\bibnamefont {Oko{\l}owicz}},\ and\ \bibinfo {author} {\bibfnamefont {M.}~\bibnamefont {P{\l}oszajczak}},\ }\bibfield  {title} {\bibinfo {title} {Theory of the two-proton radioactivity in the continuum shell model},\ }\href@noop {} {\bibfield  {journal} {\bibinfo  {journal} {Nuclear Physics A}\ }\textbf {\bibinfo {volume} {767}},\ \bibinfo {pages} {13} (\bibinfo {year} {2006})}\BibitemShut {NoStop}%
\bibitem [{\citenamefont {Wang}\ \emph {et~al.}(2026)\citenamefont {Wang}, \citenamefont {Liu},\ and\ \citenamefont {Guo}}]{Wang2026}%
  \BibitemOpen
  \bibfield  {author} {\bibinfo {author} {\bibfnamefont {Z.}~\bibnamefont {Wang}}, \bibinfo {author} {\bibfnamefont {Q.}~\bibnamefont {Liu}},\ and\ \bibinfo {author} {\bibfnamefont {J.-Y.}\ \bibnamefont {Guo}},\ }\bibfield  {title} {\bibinfo {title} {Examination of proton radioactivity in exotic nuclei with a deformed gamow-like model},\ }\href@noop {} {\bibfield  {journal} {\bibinfo  {journal} {Chinese Physics C}\ }\textbf {\bibinfo {volume} {50}},\ \bibinfo {pages} {014109} (\bibinfo {year} {2026})}\BibitemShut {NoStop}%
\bibitem [{\citenamefont {Gonçalves}\ \emph {et~al.}(2017)\citenamefont {Gonçalves}, \citenamefont {Teruya}, \citenamefont {Tavares},\ and\ \citenamefont {Duarte}}]{Gonçalves2017}%
  \BibitemOpen
  \bibfield  {author} {\bibinfo {author} {\bibfnamefont {M.}~\bibnamefont {Gonçalves}}, \bibinfo {author} {\bibfnamefont {N.}~\bibnamefont {Teruya}}, \bibinfo {author} {\bibfnamefont {O.}~\bibnamefont {Tavares}},\ and\ \bibinfo {author} {\bibfnamefont {S.}~\bibnamefont {Duarte}},\ }\bibfield  {title} {\bibinfo {title} {Two-proton emission half-lives in the effective liquid drop model},\ }\href@noop {} {\bibfield  {journal} {\bibinfo  {journal} {Physics Letters B}\ }\textbf {\bibinfo {volume} {774}},\ \bibinfo {pages} {14} (\bibinfo {year} {2017})}\BibitemShut {NoStop}%
\bibitem [{\citenamefont {Kondev}\ \emph {et~al.}(2021)\citenamefont {Kondev}, \citenamefont {Wang}, \citenamefont {Huang}, \citenamefont {Naimi},\ and\ \citenamefont {Audi}}]{Kondev2021}%
  \BibitemOpen
  \bibfield  {author} {\bibinfo {author} {\bibfnamefont {F.~G.}\ \bibnamefont {Kondev}}, \bibinfo {author} {\bibfnamefont {M.}~\bibnamefont {Wang}}, \bibinfo {author} {\bibfnamefont {W.~J.}\ \bibnamefont {Huang}}, \bibinfo {author} {\bibfnamefont {S.}~\bibnamefont {Naimi}},\ and\ \bibinfo {author} {\bibfnamefont {G.}~\bibnamefont {Audi}},\ }\bibfield  {title} {\bibinfo {title} {The {NUBASE2020} evaluation of nuclear physics properties},\ }\href@noop {} {\bibfield  {journal} {\bibinfo  {journal} {Chinese Physics C}\ }\textbf {\bibinfo {volume} {45}},\ \bibinfo {pages} {030001} (\bibinfo {year} {2021})}\BibitemShut {NoStop}%
\bibitem [{\citenamefont {Blank}\ and\ \citenamefont {Borge}(2008)}]{Blank2008}%
  \BibitemOpen
  \bibfield  {author} {\bibinfo {author} {\bibfnamefont {B.}~\bibnamefont {Blank}}\ and\ \bibinfo {author} {\bibfnamefont {M.~J.~G.}\ \bibnamefont {Borge}},\ }\bibfield  {title} {\bibinfo {title} {Nuclear structure at the proton drip line: Advances with nuclear decay studies},\ }\href@noop {} {\bibfield  {journal} {\bibinfo  {journal} {Progress in Particle and Nuclear Physics}\ }\textbf {\bibinfo {volume} {60}},\ \bibinfo {pages} {403} (\bibinfo {year} {2008})}\BibitemShut {NoStop}%
\bibitem [{\citenamefont {Doherty}\ \emph {et~al.}(2021)\citenamefont {Doherty}, \citenamefont {Andreyev}, \citenamefont {Seweryniak} \emph {et~al.}}]{Doherty2021}%
  \BibitemOpen
  \bibfield  {author} {\bibinfo {author} {\bibfnamefont {D.~T.}\ \bibnamefont {Doherty}}, \bibinfo {author} {\bibfnamefont {A.~N.}\ \bibnamefont {Andreyev}}, \bibinfo {author} {\bibfnamefont {D.}~\bibnamefont {Seweryniak}}, \emph {et~al.},\ }\bibfield  {title} {\bibinfo {title} {Solving the puzzles of the decay of the heaviest known proton-emitting nucleus {$^{185}$Bi}},\ }\href@noop {} {\bibfield  {journal} {\bibinfo  {journal} {Physical Review Letters}\ }\textbf {\bibinfo {volume} {127}},\ \bibinfo {pages} {202501} (\bibinfo {year} {2021})}\BibitemShut {NoStop}%
\bibitem [{\citenamefont {Xiao}\ \emph {et~al.}(2023)\citenamefont {Xiao}, \citenamefont {Cheng}, \citenamefont {Wang} \emph {et~al.}}]{Xiao2023}%
  \BibitemOpen
  \bibfield  {author} {\bibinfo {author} {\bibfnamefont {Q.}~\bibnamefont {Xiao}}, \bibinfo {author} {\bibfnamefont {J.-H.}\ \bibnamefont {Cheng}}, \bibinfo {author} {\bibfnamefont {B.-L.}\ \bibnamefont {Wang}}, \emph {et~al.},\ }\bibfield  {title} {\bibinfo {title} {Half-lives for proton emission and $\alpha$ decay within the deformed gamow-like model},\ }\href@noop {} {\bibfield  {journal} {\bibinfo  {journal} {Journal of Physics G: Nuclear and Particle Physics}\ }\textbf {\bibinfo {volume} {50}},\ \bibinfo {pages} {085102} (\bibinfo {year} {2023})}\BibitemShut {NoStop}%
\bibitem [{\citenamefont {Audi}\ \emph {et~al.}(2017)\citenamefont {Audi}, \citenamefont {Kondev}, \citenamefont {Wang} \emph {et~al.}}]{Audi2017}%
  \BibitemOpen
  \bibfield  {author} {\bibinfo {author} {\bibfnamefont {G.}~\bibnamefont {Audi}}, \bibinfo {author} {\bibfnamefont {F.~G.}\ \bibnamefont {Kondev}}, \bibinfo {author} {\bibfnamefont {M.}~\bibnamefont {Wang}}, \emph {et~al.},\ }\bibfield  {title} {\bibinfo {title} {The {NUBASE2016} evaluation of nuclear properties},\ }\href@noop {} {\bibfield  {journal} {\bibinfo  {journal} {Chinese Physics C}\ }\textbf {\bibinfo {volume} {41}},\ \bibinfo {pages} {030001} (\bibinfo {year} {2017})}\BibitemShut {NoStop}%
\bibitem [{\citenamefont {Wang}\ \emph {et~al.}(2021)\citenamefont {Wang}, \citenamefont {Huang}, \citenamefont {Kondev} \emph {et~al.}}]{Wang2021}%
  \BibitemOpen
  \bibfield  {author} {\bibinfo {author} {\bibfnamefont {M.}~\bibnamefont {Wang}}, \bibinfo {author} {\bibfnamefont {W.~J.}\ \bibnamefont {Huang}}, \bibinfo {author} {\bibfnamefont {F.~G.}\ \bibnamefont {Kondev}}, \emph {et~al.},\ }\bibfield  {title} {\bibinfo {title} {The {AME2020} atomic mass evaluation ({II}). tables, graphs and references},\ }\href@noop {} {\bibfield  {journal} {\bibinfo  {journal} {Chinese Physics C}\ }\textbf {\bibinfo {volume} {45}},\ \bibinfo {pages} {030003} (\bibinfo {year} {2021})}\BibitemShut {NoStop}%
\bibitem [{\citenamefont {Wang}\ \emph {et~al.}(2017)\citenamefont {Wang}, \citenamefont {Audi}, \citenamefont {Kondev} \emph {et~al.}}]{Wang2017}%
  \BibitemOpen
  \bibfield  {author} {\bibinfo {author} {\bibfnamefont {M.}~\bibnamefont {Wang}}, \bibinfo {author} {\bibfnamefont {G.}~\bibnamefont {Audi}}, \bibinfo {author} {\bibfnamefont {F.~G.}\ \bibnamefont {Kondev}}, \emph {et~al.},\ }\bibfield  {title} {\bibinfo {title} {The {AME2016} atomic mass evaluation ({II}). tables, graphs and references},\ }\href@noop {} {\bibfield  {journal} {\bibinfo  {journal} {Chinese Physics C}\ }\textbf {\bibinfo {volume} {41}},\ \bibinfo {pages} {030003} (\bibinfo {year} {2017})}\BibitemShut {NoStop}%
\bibitem [{\citenamefont {Ma}\ \emph {et~al.}(2019)\citenamefont {Ma}, \citenamefont {Zhang}, \citenamefont {Bao},\ and\ \citenamefont {Zhang}}]{Ma2019}%
  \BibitemOpen
  \bibfield  {author} {\bibinfo {author} {\bibfnamefont {N.-N.}\ \bibnamefont {Ma}}, \bibinfo {author} {\bibfnamefont {H.-F.}\ \bibnamefont {Zhang}}, \bibinfo {author} {\bibfnamefont {X.-J.}\ \bibnamefont {Bao}},\ and\ \bibinfo {author} {\bibfnamefont {H.-F.}\ \bibnamefont {Zhang}},\ }\bibfield  {title} {\bibinfo {title} {Basic characteristics of nuclear landscape by improved weizsäcker-skyrme-type nuclear mass model *},\ }\href@noop {} {\bibfield  {journal} {\bibinfo  {journal} {Chinese Physics C}\ }\textbf {\bibinfo {volume} {43}},\ \bibinfo {pages} {044105} (\bibinfo {year} {2019})}\BibitemShut {NoStop}%
\bibitem [{\citenamefont {Liu}\ \emph {et~al.}(2024)\citenamefont {Liu}, \citenamefont {Jiang}, \citenamefont {Wu} \emph {et~al.}}]{Liu2024}%
  \BibitemOpen
  \bibfield  {author} {\bibinfo {author} {\bibfnamefont {X.}~\bibnamefont {Liu}}, \bibinfo {author} {\bibfnamefont {J.-D.}\ \bibnamefont {Jiang}}, \bibinfo {author} {\bibfnamefont {X.-J.}\ \bibnamefont {Wu}}, \emph {et~al.},\ }\bibfield  {title} {\bibinfo {title} {Systematic study of cluster radioactivity in trans-lead nuclei with various versions of proximity potential formalisms},\ }\href@noop {} {\bibfield  {journal} {\bibinfo  {journal} {Chinese Physics C}\ }\textbf {\bibinfo {volume} {48}},\ \bibinfo {pages} {054101} (\bibinfo {year} {2024})}\BibitemShut {NoStop}%
\bibitem [{\citenamefont {Zhang}\ \emph {et~al.}(2009)\citenamefont {Zhang}, \citenamefont {Dong}, \citenamefont {Royer} \emph {et~al.}}]{Zhang2009}%
  \BibitemOpen
  \bibfield  {author} {\bibinfo {author} {\bibfnamefont {H.~F.}\ \bibnamefont {Zhang}}, \bibinfo {author} {\bibfnamefont {J.~M.}\ \bibnamefont {Dong}}, \bibinfo {author} {\bibfnamefont {G.}~\bibnamefont {Royer}}, \emph {et~al.},\ }\bibfield  {title} {\bibinfo {title} {Preformation of clusters in heavy nuclei and cluster radioactivity},\ }\href@noop {} {\bibfield  {journal} {\bibinfo  {journal} {Physical Review C}\ }\textbf {\bibinfo {volume} {80}},\ \bibinfo {pages} {037307} (\bibinfo {year} {2009})}\BibitemShut {NoStop}%
\bibitem [{\citenamefont {Warda}\ and\ \citenamefont {Robledo}(2011)}]{Warda2011}%
  \BibitemOpen
  \bibfield  {author} {\bibinfo {author} {\bibfnamefont {M.}~\bibnamefont {Warda}}\ and\ \bibinfo {author} {\bibfnamefont {L.~M.}\ \bibnamefont {Robledo}},\ }\bibfield  {title} {\bibinfo {title} {Microscopic description of cluster radioactivity in actinide nuclei},\ }\href@noop {} {\bibfield  {journal} {\bibinfo  {journal} {Physical Review C}\ }\textbf {\bibinfo {volume} {84}},\ \bibinfo {pages} {044608} (\bibinfo {year} {2011})}\BibitemShut {NoStop}%
\bibitem [{\citenamefont {Saidi}\ \emph {et~al.}(2015)\citenamefont {Saidi}, \citenamefont {Oudih}, \citenamefont {Fellah} \emph {et~al.}}]{Saidi2015}%
  \BibitemOpen
  \bibfield  {author} {\bibinfo {author} {\bibfnamefont {F.}~\bibnamefont {Saidi}}, \bibinfo {author} {\bibfnamefont {M.~R.}\ \bibnamefont {Oudih}}, \bibinfo {author} {\bibfnamefont {M.}~\bibnamefont {Fellah}}, \emph {et~al.},\ }\bibfield  {title} {\bibinfo {title} {Cluster decay investigation within a modified woods--saxon potential},\ }\href@noop {} {\bibfield  {journal} {\bibinfo  {journal} {Modern Physics Letters A}\ }\textbf {\bibinfo {volume} {30}},\ \bibinfo {pages} {1550150} (\bibinfo {year} {2015})}\BibitemShut {NoStop}%
\bibitem [{\citenamefont {Liao}\ \emph {et~al.}(2025)\citenamefont {Liao}, \citenamefont {Fan}, \citenamefont {Liu} \emph {et~al.}}]{Liao2025}%
  \BibitemOpen
  \bibfield  {author} {\bibinfo {author} {\bibfnamefont {L.-J.}\ \bibnamefont {Liao}}, \bibinfo {author} {\bibfnamefont {Y.}~\bibnamefont {Fan}}, \bibinfo {author} {\bibfnamefont {X.}~\bibnamefont {Liu}}, \emph {et~al.},\ }\bibfield  {title} {\bibinfo {title} {Laser-assisted cluster radioactivity within a deformed gamow-like model},\ }\href@noop {} {\bibfield  {journal} {\bibinfo  {journal} {Physical Review C}\ }\textbf {\bibinfo {volume} {112}},\ \bibinfo {pages} {014327} (\bibinfo {year} {2025})}\BibitemShut {NoStop}%
\end{thebibliography}%

\end{document}